\documentclass[12pt]{article}

\usepackage{epsf}
\usepackage{graphicx}
\usepackage{axodraw}
\usepackage{cite}
\usepackage{amsmath}

\newcommand\lsim{\mathrel{\rlap{\lower4pt\hbox{\hskip1pt$\sim$}}
    \raise1pt\hbox{$<$}}}
\newcommand\gsim{\mathrel{\rlap{\lower4pt\hbox{\hskip1pt$\sim$}}
    \raise1pt\hbox{$>$}}}
\def\lsim{\mathrel{\raise.3ex\hbox{$<$\kern-.75em\lower1ex\hbox{$\sim$}}}} 
\def\gsim{\mathrel{\raise.3ex\hbox{$>$\kern-.75em\lower1ex\hbox{$\sim$}}}}

\def\beq{\begin{equation}}
\def\eeq{\end{equation}}
\def\ba{\begin{eqnarray}}
\def\ea{\end{eqnarray}}

\begin{document}
\begin{titlepage}
\pagestyle{empty}
\baselineskip=21pt

\vskip -4in
{\rightline{\small 
%CERN-PH-TH/2010-014, SLAC-PUB-xxxxx, UFIFT-HEP-10-xx}}
%CERN-PH-TH/2010-014, SLAC-PUB-xxxxx}}
CERN-PH-TH/2010-014}}
\vskip 0.3in
\begin{center}
{\Large {Extra Dimensions at the LHC}}
\end{center}
\begin{center}
\vskip 0.05in
{{Kyoungchul Kong}~$^a$, 
{Konstantin Matchev}~$^b$ and 
{G\'eraldine Servant}~$^{c,d}$ }\\
\vskip 0.2in
{\it \small
$^a$ Theoretical Physics Department, SLAC, Menlo Park, CA 94025, USA \\
$^b$ Physics Department, University of Florida, Gainesville, FL 32611, USA \\
$^c$ CERN Physics Department, Theory Division, CH-1211 Geneva 23, Switzerland \\ 
$^d$ Institut de Physique Th\'eorique, CEA Saclay, F91191 Gif-sur-Yvette, France \\}
\vskip 0.4in
{\bf Abstract}
\end{center}
\baselineskip=18pt \noindent
%%%%%%%%%%%%%%%%%%%%%%%%%%%%%%%%%%%%%%%%%%%%%%%

{\small 
\vspace{0.5cm}
We discuss the motivation and the phenomenology of models with 
either flat or warped extra dimensions. We describe the typical mass 
spectrum and discovery signatures of such models at the LHC.
We also review several proposed methods for discriminating the usual 
low-energy supersymmetry from a model with flat (universal) extra 
dimensions. 
\vspace{4cm}
}

\begin{center}
{\it \small From  `Particle Dark Matter: Observations, Models and Searches' \\
edited by Gianfranco Bertone \\
Copyright $@$ 2010 Cambridge University Press. \\
Chapter 15, pp. 306-324, Hardback  ISBN  9780521763684, 
http://cambridge.org/us/catalogue/catalogue.asp?isbn=9780521763684}
\end{center}

%%%%%%%%%%%%%%%%%%%%%%%%%%%%%%%%%%%%%%%%%%%%%%%%%%%

%\vskip 0.15in

%\leftline{\today}
\end{titlepage}
\baselineskip=18pt

%%%%%%%%%%%%%%%%%%%%%%%%%%%%%%%%%%%%%%%%%%%%%%%%%%%

% cup2egui.tex (LaTeX 2e version)
% v1.01 --- released 9th July 1997
% v1.0  --- released 9th May 1997
%           based on cupguide.tex v1.2 (for LaTeX2.09, 27.4.95)

%\NeedsTeXFormat{LaTeX2e}[1996/06/01]

%\documentclass[cup6b]{cupbook}
%\setcounter{tocdepth}{4}
%\usepackage{natbib}
%\usepackage{natpatch}

%\title[Particle DM]
%      {Particle Dark Matter}
%\author{Gianfranco Bertone}
%\date{today}

\section{Extra Dimensions at the LHC}

In models with extra dimensions, the usual $3+1$ dimensional space-time 
$x^\mu \equiv (x^0,x^1,x^2,x^3)$ is extended to include additional
spatial dimensions parameterized by coordinates $x^4,x^5,\ldots,x^{3+N}$.
Here $N$ is the number of extra dimensions. String theory arguments
would suggest that in principle $N$ can be as large as 6 or 7.
In this chapter, we are interested in extra dimensional (ED) models
where all particles of the Standard Model (SM)
are allowed to propagate in the bulk, i.e. along any of the 
$x^{3+i}$ $(i=1,\ldots,N)$ directions \cite{Appelquist:2000nn}. 
In order to avoid a blatant contradiction with the observed reality,
the extra dimensions in such models must be extremely small:
smaller than the smallest scale which has been currently resolved 
by experiment. Therefore, the extra dimensions are assumed to be suitably 
compactified on some manifold of sufficiently small size 
(see Fig.~\ref{fig:UED}).

Depending on the type of metric in the bulk, 
the ED models fall into one of the following two 
categories: flat, a.k.a.~``universal'' extra dimensions (UED) 
models, discussed in Section~\ref{sec:flatUED},
or warped ED models, discussed in Section~\ref{sec:warpedUED}.
As it turns out, the collider signals of the ED models
are strikingly similar to the signatures of supersymmetry (SUSY) \cite{Cheng:2002ab}.
%discussed in Ref. \cite{Cheng:2002ab}. % Chapter \ref{Chap:Plehn} \cite{Cheng:2002ab}. 
Section \ref{sec:susyUED} outlines some general methods for 
distinguishing an ED model from SUSY at high energy colliders.

\section{Flat extra dimensions (UED)}
\label{sec:flatUED}

\subsection{Definition}
\label{sec:flatUEDdef}

\begin{figure}[tb]
\unitlength=1.5 pt
\SetScale{1.5}
\SetWidth{1.0}      % line    size control
\normalsize    %  letter  size control
{} \qquad\allowbreak
%  diagram # 1
\centerline{
\begin{picture}(260,130)(20,10)
\SetColor{Gray}
\LongArrow( 75, 75)( 75, 45)
\LongArrow( 75, 75)( 75,105)
\LongArrow( 57, 75)( 57, 51)
\LongArrow( 57, 75)( 57, 99)
\LongArrow( 93, 75)( 93, 51)
\LongArrow( 93, 75)( 93, 99)
\LongArrowArc(75,75)(42,0,40)
\LongArrow(154, 40)(154, 70)
\LongArrow(160, 34)(190, 34)
\LongArrowArc(160,40)(25,45,85)
\LongArrowArc(160,40)(45,45,87)
\LongArrowArcn(160,40)(25,45,5)
\LongArrowArcn(160,40)(45,45,3)
\LongArrowArc(230,110)(25,225,265)
\LongArrowArc(230,110)(45,225,267)
\LongArrowArcn(230,110)(25,225,185)
\LongArrowArcn(230,110)(45,225,183)
\Text(122,90)[c]{\Black{$x^4$}}
\Text(175,27)[c]{\Black{$x^4$}}
\Text(147,55)[c]{\Black{$x^5$}}
\Text( 75,15)[c]{(a)}
\Text(195,15)[c]{(b)}
\SetColor{Red}
\Line(160, 40)(230, 40)
\Line(160,110)(230,110)
\Line(160, 40)(160,110)
\Line(230, 40)(230,110)
\CArc( 75, 75)(35,0,360)
%\Text( 75, 86)[c]{\Red{$D$}}
%\Text(125, 86)[c]{\Red{$C$}}
%\Text(175, 86)[c]{\Red{$B$}}
%\Text(225, 86)[c]{\Red{$A$}}
%\SetWidth{1.2}      % line    size control
%\Line(50,80)(250,80)
%\Text(100, 55)[c]{\Black{$x=\frac{m_C^2}{m_D^2}$}}
%\Text(150, 55)[c]{\Black{$y=\frac{m_B^2}{m_C^2}$}}
%\Text(200, 55)[c]{\Black{$z=\frac{m_A^2}{m_B^2}$}}
%\Text(120,129)[c]{\Black{$q$}}
%\Text(170,129)[c]{\Black{$\ell_n$}}
%\Text(220,129)[c]{\Black{$\ell_f$}}
\SetColor{Blue}
\Vertex( 40, 75){2}
\Vertex(110, 75){2}
\Vertex(160, 40){2}
\Vertex(230,110){2}
\Vertex(160,110){2}
\Vertex(230, 40){2}
\LongArrow(195, 75)(225, 45)
\LongArrow(195, 75)(165,105)
%\scriptsize
%\Text(100, 73)[c]{\Blue{$c_LP_L+c_RP_R$}}
%\Text(150, 73)[c]{\Blue{$b_LP_L+b_RP_R$}}
%\Text(200, 73)[c]{\Blue{$a_LP_L+a_RP_R$}}
\end{picture} 
}
\caption{(a) Compactification of $N=1$ extra dimension
on a circle with opposite points identified. 
(b) Compactification of $N=2$ extra dimensions
on the chiral square with adjacent sides identified.
In each case, the black arrows indicate the 
corresponding identification.
The blue dots represent fixed (boundary) points.}
\label{fig:UED} 
\end{figure}
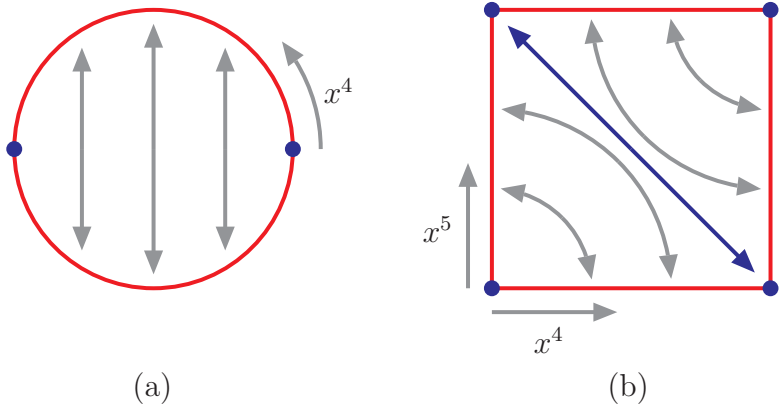

In this section, we choose the metric on the extra dimensions to be flat.
For simplicity, we shall limit our discussion to models 
with $N=1$ or $N=2$ UEDs. In the simplest case of $N=1$, a
compact extra dimension $x^4$ would have the topology of a 
circle $S^1$. However, in order to implement the chiral fermions
of the SM in a UED framework, one must use a manifold with endpoints,
e.g.~the orbifold $S^1/Z_2$, pictorially represented in 
Fig.~\ref{fig:UED}(a). The opposite sides of the circle are identified, 
as indicated by the black vertical arrows. The net result is a
single line segment with two endpoints, denoted by the blue dots. 
The size of the extra dimension in this case is simply parameterized 
by the radius of the circle $R$.

%%%%%%%%%%%%%%%%%%% BEGIN FIGURE %%%%%%%%%%%%%%%%%%%
%\begin{center}

%\end{center}
%%%%%%%%%%%%% END OF FIGURE ################

In the case of two extra dimensions ($N=2$) there are several 
possibilities for compactification. One of them is the so called
``chiral square'' and corresponds to a $T^2/Z_4$ orbifold 
\cite{Dobrescu:2004zi}. It can be visualized as shown in 
Fig.~\ref{fig:UED}(b). The two extra dimensions have equal 
size, and the boundary conditions are such that 
the adjacent sides of the ``chiral square'' are identified, 
as indicated by the black arrows. The resulting orbifold 
endpoints are again denoted by blue dots.

An important concept in any UED model is the notion of 
Kaluza-Klein (KK) parity, whose origin can be traced back to
the geometrical symmetries of the compactification.
For example, in the $N=1$ case of Fig.~\ref{fig:UED}(a),
KK-parity corresponds to the reflection symmetry with respect 
to the center of the line segment. Similarly, in the 
$N=2$ case of Fig.~\ref{fig:UED}(b), KK parity is due to
the symmetry with respect to the center of the chiral 
square\footnote{Notice that there is only one KK parity since 
the two directions $x^4$ and $x^5$ of the chiral square 
are related to each other through the boundary condition.}.
In general, UED models respect KK parity, and this fact has important
consequences for their collider and astroparticle phenomenology.

\subsection{Mass spectrum}
\label{sec:flatUEDmass}

Since the extra dimensions are compact, the extra-dimensional components of 
the momentum of any SM particle are quantized in units of $\frac{1}{R}$. 
From the usual 4-dimensional point of view, those 
momentum components are interpreted as masses.
Therefore in UED models each SM particle is accompanied by an 
infinite tower of heavy KK particles with masses $\frac{n}{R}$,
where the integer $n$ counts the number of quantum units of 
extra-dimensional momentum. All KK particles at a given $n$ 
are said to belong to the $n$-th KK level, and at leading 
order appear to be exactly degenerate.

\begin{figure}
\centerline{
\includegraphics[width=6.5cm]{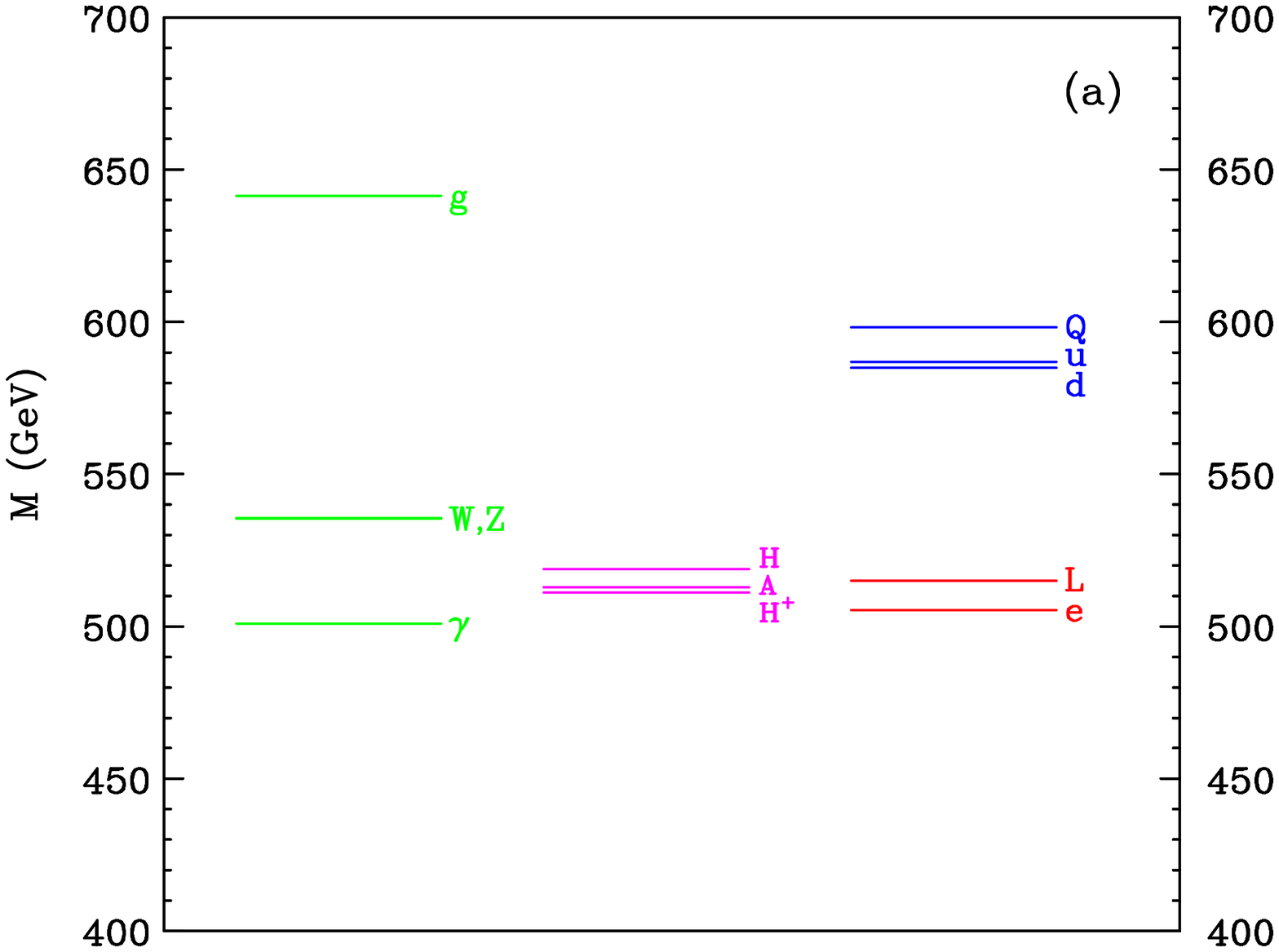}\hspace*{0.4cm}
\includegraphics[width=6.5cm]{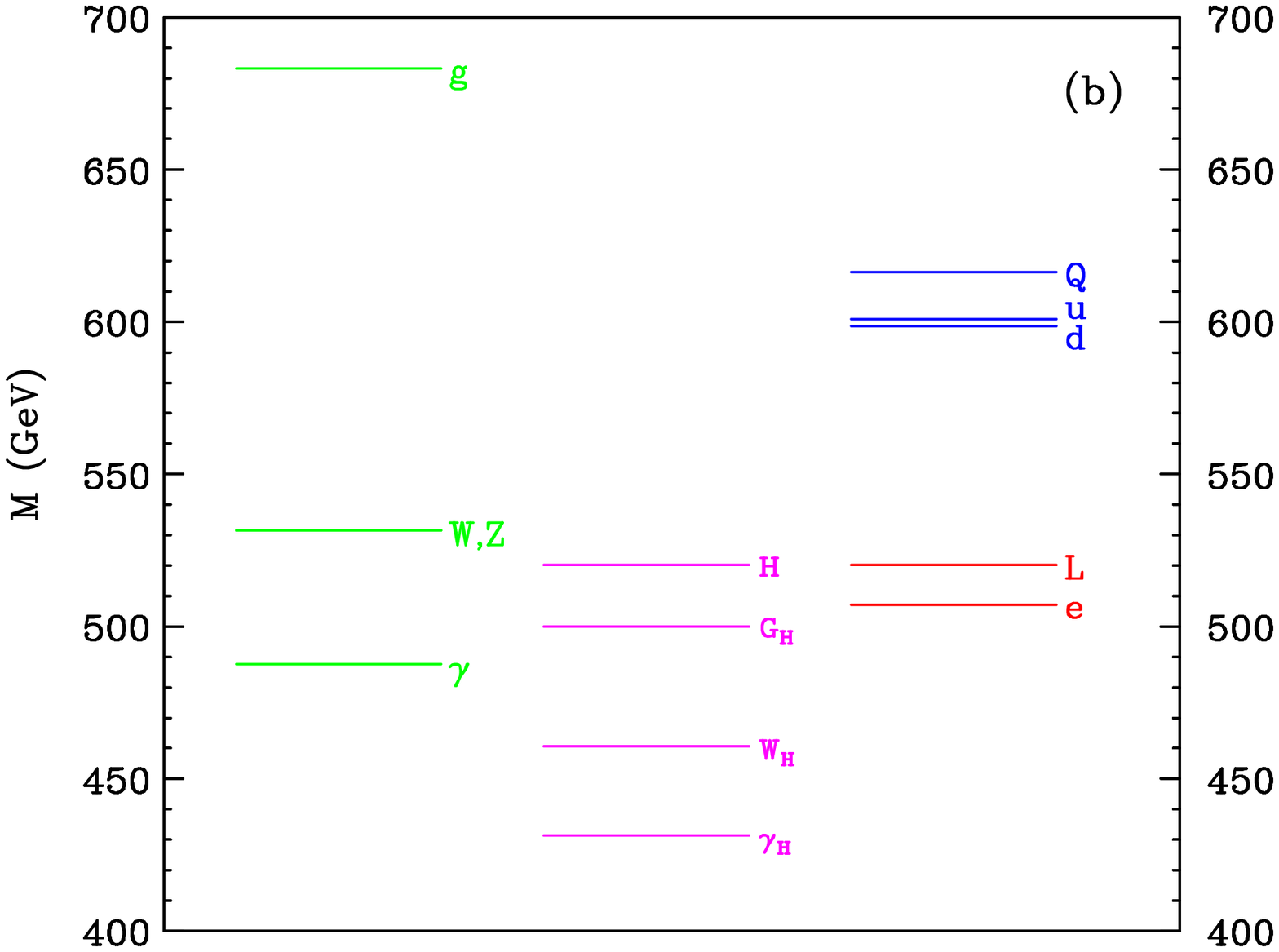} }
\caption{The radiatively corrected mass spectrum of the 
level one ($n=1$) KK particles in the two UED models 
from Fig.~\ref{fig:UED}, for $R^{-1}=500$ GeV.
In each panel, from left to right we list the KK particles 
of spin-1 (green), spin-0 (magenta) and spin-$\frac{1}{2}$
(blue for KK quarks and red for KK leptons). 
Figures taken from Refs. \cite{Cheng:2002iz} and \cite{Burdman:2006gy}.}
\label{fig:mass}
\end{figure}

However, the masses of the KK particles receive corrections from several 
sources, which will lift this degeneracy. First, there are 
tree-level corrections arising from electroweak symmetry breaking
through the usual Higgs mechanism. More importantly, there are
one-loop mass renormalization effects due to the usual SM interactions
in the bulk \cite{Cheng:2002iz}. Finally, there may be contributions from 
boundary terms which live on the orbifold fixed points (the blue dots
in Fig.~\ref{fig:UED}) \cite{Cheng:2002iz,Flacke:2008ne}. 
In the so-called ``minimal UED'' models, 
the last effect is ignored and the resulting one-loop radiatively corrected 
spectrum of the first level KK modes is as shown in Fig.~\ref{fig:mass}.

In the case of $N=1$ minimal UED shown in Fig.~\ref{fig:mass}(a),
we find that the SM particle content is simply duplicated 
at the $n=1$ level. The mass splittings among the KK particles arise 
mainly due to radiative corrections, which are largest for 
the colored particles (KK quarks and KK gluon). 
The lightest KK particle (LKP) at level one in this case
is denoted by $\gamma_1$ and represents a linear superposition 
of the KK modes of the hypercharge gauge boson $B_1$ and
the neutral component of the $SU(2)$ gauge boson $W^0_1$.

Fig.~\ref{fig:mass}(b) reveals that the $n=1$ KK mass spectrum 
is somewhat more complex in the case of 2 extra dimensions ($N=2$).
This is because gauge bosons propagating in 5+1 dimensions may 
be polarized along either of the two extra dimensions.
As a result, for each spin-1 KK particle associated with 
a gauge boson, there are two spin-0 fields transforming in 
the adjoint representation of the gauge group. One linear
combination of those becomes the longitudinal degree of freedom 
of the spin-1 KK particle, while the other linear combination 
remains as a physical spin-0 particle, called the spinless adjoint.
In Fig.~\ref{fig:mass}(b) the spinless adjoints are 
designated by an index ``H''. Fig.~\ref{fig:mass}(b) also reveals
that in the minimal $N=2$ UED model, the LKP is the spinless 
photon $\gamma_{\rm H}$ \cite{Ponton:2005kx,Burdman:2006gy}.

\subsection{Collider signals}
\label{sec:flatUEDcol}

In terms of the KK level number\footnote{In the case of $N=2$, each 
KK particle is characterized by two indices, $n_1$ and $n_2$, 
counting the quantum units of momentum along each extra dimension.
The KK parity is then defined in terms of the total KK level number
$n_1+n_2$ as $(-1)^{n_1+n_2}$.} $n$, KK parity can be simply
defined as $(-1)^{n}$. The usual Standard Model particles do not
have any extra-dimensional momentum, and therefore have $n=0$ and 
positive KK parity. On the other hand, the KK particles can have
either positive or negative KK parity, depending on the value of $n$. 
Notice that the lightest KK particle at $n=1$ (i.e.~the LKP) 
has negative KK parity and is absolutely stable, since KK parity 
conservation prevents it from decaying into SM particles.
The collider phenomenology of the UED models is 
therefore largely determined by the nature of the LKP.
In both of the minimal UED models shown in Fig.~\ref{fig:mass},
the LKP is a neutral weakly interacting particle, 
whose signature will be missing energy in the detector. 
At hadron colliders the total parton level energy in the 
collision is a priori unknown, hence the presence of LKP particles
in the event must be inferred from an imbalance in the total 
transverse momentum. 

\begin{figure}
\centerline{
\includegraphics[width=6.5cm]{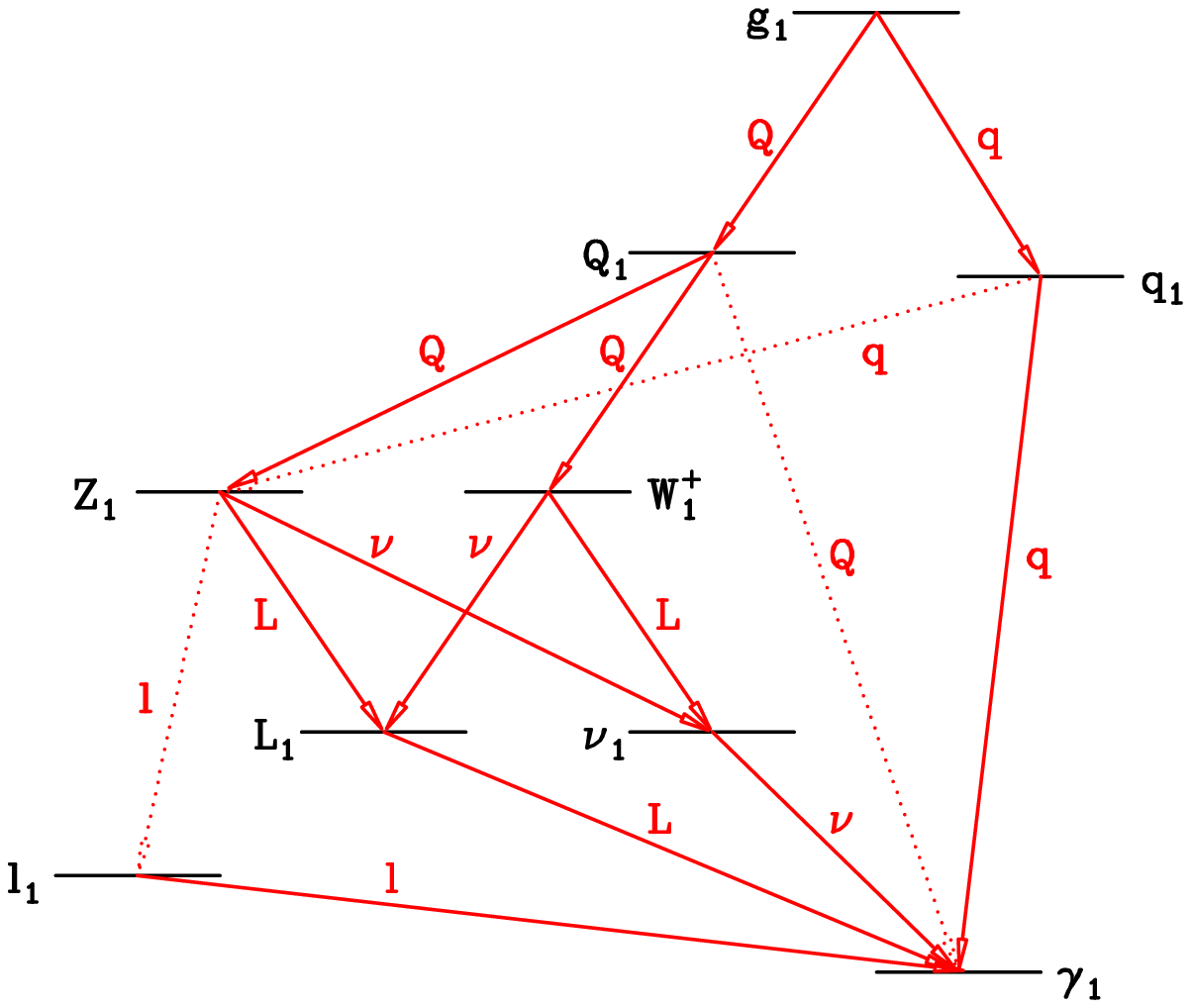}
\includegraphics[width=6.5cm]{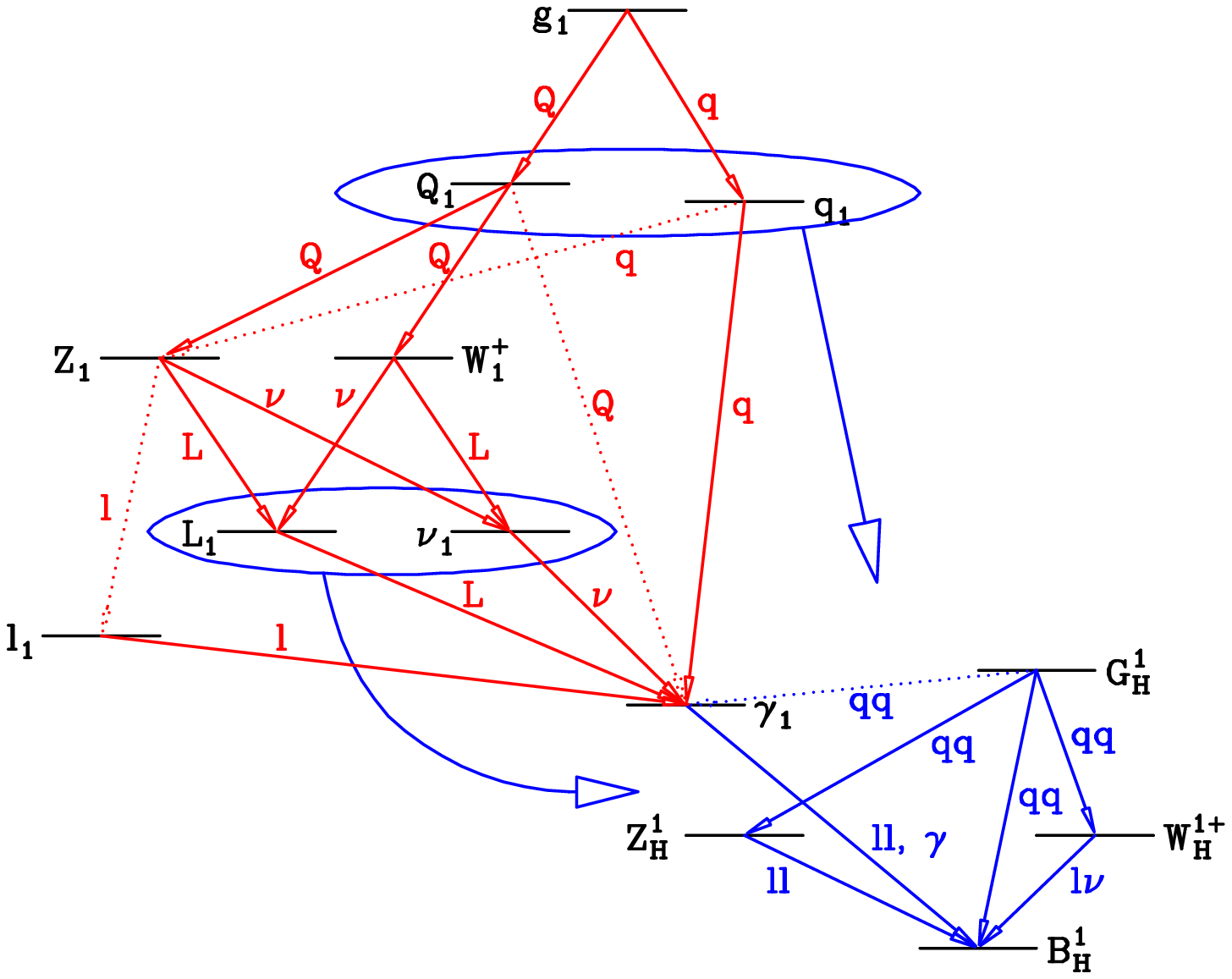} }
\caption{Qualitative sketch of the level 1 KK spectroscopy depicting
the dominant (solid) and rare (dotted) transitions and
the corresponding decay product for the $N=1$ (left) and $N=2$ (right) 
UED models from Fig.~\ref{fig:mass}.
Figures taken from Refs. \cite{Cheng:2002ab} and \cite{Freitas:2007rh}.}
\label{fig:signatures}
\end{figure}

The collider phenomenology of the minimal UED models 
from Fig.~\ref{fig:mass}
has been extensively investigated at both linear colliders 
\cite{Battaglia:2005zf,Bhattacherjee:2005qe,Riemann:2005es,%
Battaglia:2005ma,Freitas:2007rh,Ghosh:2008ji,Ghosh:2008dp}
and hadron colliders \cite{Rizzo:2001sd,Macesanu:2002db,Cheng:2002ab,Datta:2005zs,%
Burdman:2006gy,Dobrescu:2007xf,Ghosh:2008ji,Ghosh:2008ix}.
Due to KK parity conservation, the KK particles are always pair-produced, 
and then each one undergoes a cascade decay to the LKP, as
illustrated in Fig.~\ref{fig:signatures} \cite{Cheng:2002ab,Dobrescu:2007xf}. 
It is interesting to 
notice that the decay patterns in UED look very similar to those 
arising in R-parity conserving supersymmetry.
% (see Chapter \ref{Chap:Plehn}). 
The typical UED signatures include a certain number of jets, 
a certain number of leptons and photons, plus missing energy $\not \!\! E_T$. 

\begin{figure}
\centerline{
\includegraphics[width=7.5cm]{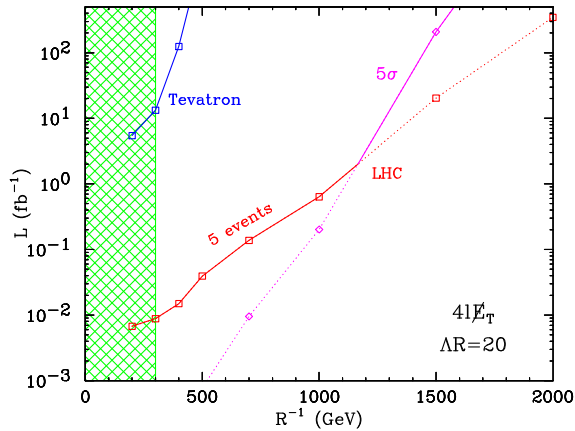} }
\caption{Discovery reach for the minimal $N=1$ UED model at the Tevatron (blue)
and the LHC (red) in the $4\ell + \not \!\! E_T$ channel. The plot shows the total
integrated luminosity per experiment, which is required for a $5\sigma$ 
discovery, given the observation of at least 5 events.
Figure taken from Ref. \cite{Cheng:2002ab}.}
\label{fig:reach4l}
\end{figure}

Which particular signature among all these offers the best prospects for discovery?
The answer to this question depends on the interplay between the 
predicted signal rates in UED and the expected SM backgrounds.
For example, at lepton colliders the SM backgrounds are firmly 
under control, and the best channel is typically the one with 
the largest signal rate. Since at lepton colliders the new
particles are produced through the same (electroweak) interactions, 
the largest rates are associated with the lightest particles 
in the spectrum - the KK leptons and the electroweak KK gauge bosons.
In contrast, at hadron colliders the dominant production is through
strong interactions, and the largest cross-sections belong to the 
colored KK particles, which typically decay through jets.
Unfortunately, the SM QCD backgrounds to the 
jetty UED signatures are significant, and in that case
there is a substantial benefit in looking for leptons instead.
As seen in Fig.~\ref{fig:signatures}, the decays of the 
weakly interacting KK quarks $Q_1$ proceed through $SU(2)$
KK gauge bosons $W^\pm_1$ and $Z_1$, whose decays are often 
accompanied by leptons (electrons or muons). The inclusive pair 
production of a $Q_1\bar{Q}_1$ pair therefore may yield up to 4 leptons, 
plus missing energy. Fig.~\ref{fig:reach4l} shows the corresponding
discovery reach for the minimal $N=1$ UED model at the Tevatron (blue) 
and the LHC (red) in the $4\ell + \not \!\! E_T$ channel \cite{Cheng:2002ab}. Recently
the CDF collaboration performed a search for the minimal $N=1$ UED model
in a multi-lepton channel, based on 100 pb$^{-1}$ of data at $\sqrt{s}$= 1.8 TeV.
That analysis placed a lower limit on the UED scale 
$R^{-1}$ of 280 GeV (at 95\% C.L.) \cite{Lin:2005ix}.

From Fig.~\ref{fig:mass}(b) one can notice that 
due to the presence of additional KK particles (spinless adjoints) 
at each level, the cascade decays become longer and may yield 
events with even more leptons. Such events with very high lepton multiplicity
would be a smoking gun for the $N=2$ UED model \cite{Dobrescu:2007xf}. 

%%%%%%%%%%%%%%%%%%%%%%%%%%%
\section{Warped extra dimensions}
\label{sec:warpedUED}

The UED models discussed in the previous section are very peculiar:
%In the framework of UED, the existence of KK parity requires very special conditions. 
the extra dimension is an interval with a flat background geometry, and
KK parity is realized as a geometric reflection about the midpoint of the extra dimension.
It is important to note that
KK parity has a larger parent symmetry, 
KK number conservation, which is broken only by the interactions living 
on the orbifold boundary points (the blue dots in Fig.~\ref{fig:UED}).
%on the branes at the ends of the interval. 
In the literature on UED models, it is usually assumed that the boundary
interactions are symmetric with respect to the $Z_2$ reflection about the midpoint, so that KK parity is an exact symmetry. It is also assumed that they are
suppressed (loop-induced), implying that KK number is still an approximate symmetry.  These assumptions have very important phenomenological implications, as both KK parity and the approximate
KK number conservation are needed to evade precision electroweak constraints for UED models. KK parity eliminates  couplings of a single odd KK mode with the SM fields, whereas the approximate 
KK number conservation
suppresses certain interactions among the even level KK modes, such as single coupling of the second KK mode with 
the SM, which are not
forbidden by KK parity. In the end, both the odd and even KK modes are allowed to have masses well
below 1 TeV. If there were only KK parity and not the approximate KK number conservation, experimental
constraints would have required the second KK mass to be higher than 2 - 3 TeV and, therefore, the compactification
scale to be around 1 TeV or higher (recall that in flat geometry KK modes are evenly spaced).

The flatness of wave function profiles in UED is not natural and reflects the fact that electroweak 
symmetry breaking is not addressed but just postulated. A model of dynamical symmetry breaking in UED would typically 
spoil the flatness of the Higgs profile and constraints on the KK scale would have to be reexamined 
accordingly. The virtue of UED is that mass scales of new particles are allowed to be very close to the 
electroweak scale at a few hundreds GeV, allowing for easy access at the LHC, and offering an interesting benchmark model for LHC searches. 
However, the UED model does not address the hierarchy between the Planck and the weak scale, 
nor does it address the fermion mass hierarchy. 
In contrast, as shown by Randall and Sundrum  \cite{Randall:1999ee}, warped extra dimensions have provided a new framework for addressing the hierarchy problem in extensions of the Standard Model.

\subsection{Generic features of warped spacetime}

\begin{figure}[tb]
\begin{center}
\includegraphics[height=8cm]{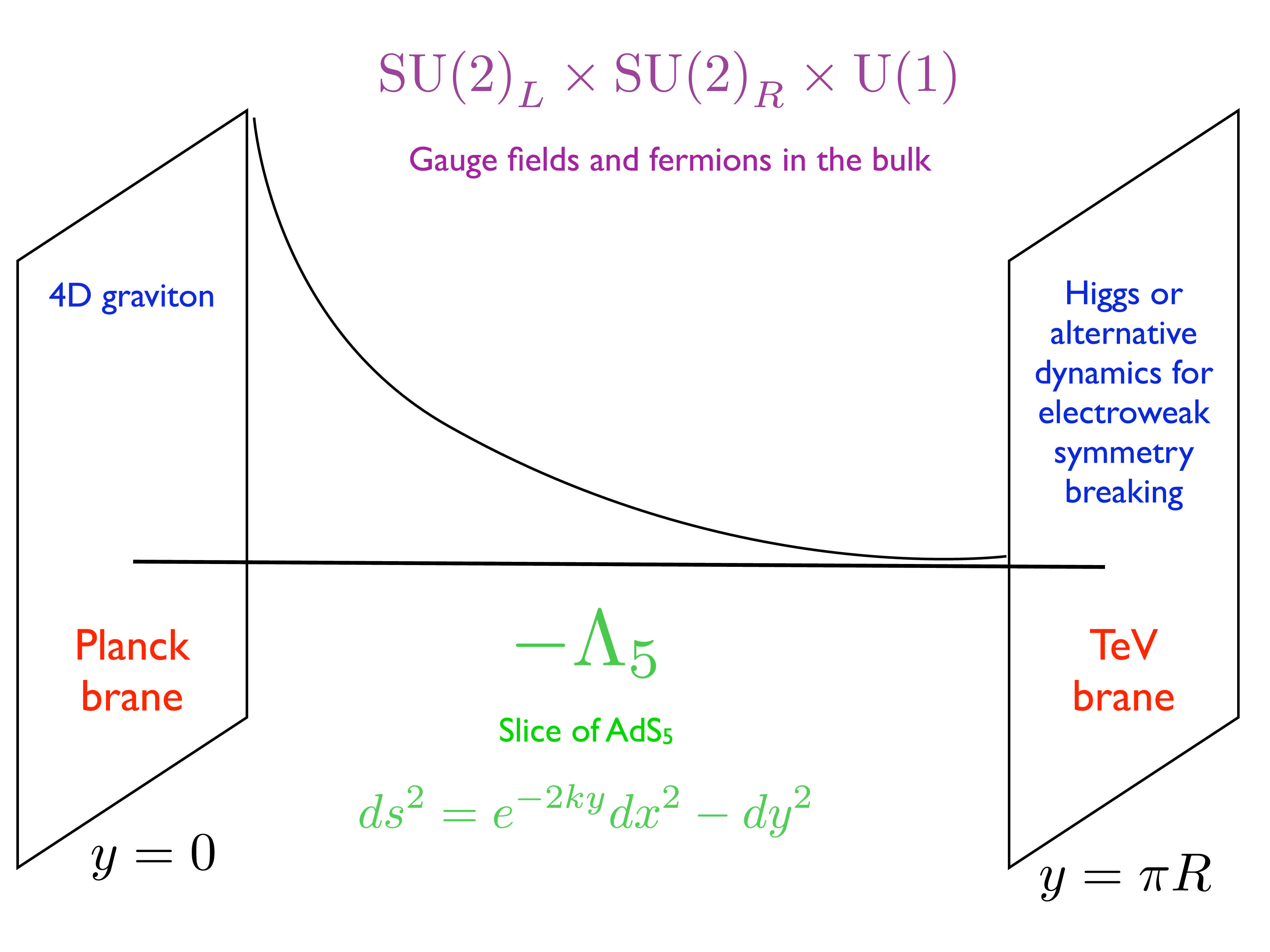}
\end{center}
\caption[]{Randall--Sundrum set-up.}
\label{fig:RSgeometry}
\end{figure}

The  Randall--Sundrum (RS) solution \cite{Randall:1999ee} is based on 
a slice of five-dimensional anti de Sitter space AdS$_5$ bounded by two three-branes, the UV and IR branes.  The background space time metric  of AdS$_5$ is
\begin{equation}
ds^2=e^{-2ky}\eta_{\mu\nu}dx^{\mu}dx^{\nu}-dy^2 \, ,
\end{equation}
where $k$ is the AdS curvature scale of order the Planck scale (fixed by the bulk cosmological constant). The $y$ dependence in the metric is known as the ``warp" factor. The UV (IR) brane is located at $y=0$ ($y=\pi R$) (see Fig.~\ref{fig:RSgeometry}).
The key point is that distance scales are measured with the nonfactorisable metric of AdS space. Hence,  energy scales are location dependent  along the fifth  dimension and the hierarchy problem can be redshifted away.
The RS model supposes that all fundamental mass parameters are of order the Planck scale.
Owing to the warped geometry, the 4D (or zero-mode) graviton is localized near the UV/Planck brane, 
whereas the Higgs sector can be localized near the IR brane where the cut-off scale is scaled down to $M_{\tiny{Planck}} e^{-\pi k R}$. 
According to the AdS/CFT correspondence \cite{Maldacena:1997re, Witten:1998qj}, AdS$_5$  is  dual to a 4D strongly-coupled conformal field theory (CFT). Thus, the RS solution is conjectured to be dual to composite Higgs models \cite{ArkaniHamed:2000ds,Rattazzi:2000hs,Contino:2003ve} where the TeV scale is hierarchically smaller and stable compared to the UV scale.

In the original RS model, the entire SM was assumed to be localized on the TeV brane.
It was subsequently realized
that when the SM fermion \cite{Grossman:1999ra, Gherghetta:2000qt} and gauge fields
\cite{Davoudiasl:1999tf,Pomarol:1999ad,Chang:1999nh} are allowed to propagate in the bulk,
such a framework not only solves the Planck-weak hierarchy, but can 
also address the flavor hierarchy.
The idea is that light SM fermions
(which are zero-modes of 5D fermions)  can be localized near the
UV brane, whereas the top quark is localized near the
IR brane, resulting in small and large couplings respectively
to the SM Higgs localized near the IR brane.
In the CFT language, the Standard Model fermions and 
gauge bosons are partly composite to varying degrees, ranging from an elementary 
electron to a composite top quark. 
%
%Moreover, the flavor problem 
%(both from unknown physics at the cut-off and from the KK states)
%is also under control
%\cite{gp, Huber:2000ie} due to an analog of the GIM mechanism  or
%approximate
%flavor symmetries \cite{Agashe:2004cp},
%even with a few TeV KK scale and despite
 %the recent $B$-physics data \cite{NMFV}.
Finally, a central requirement in these constructions is having an approximate ``custodial isospin" symmetry of the strong sector to protect the EW  $\rho$ parameter. This is ensured by extending the EW gauge group to $SU(2)_L \times SU(2)_R \times U(1)  $ \cite{Agashe:2003zs}.

In the RS setup,  the KK modes are generically  localized towards the IR brane. On the other hand,
the zero mode gauge bosons have a flat profile along the extra dimension while the zero mode fermions can be arbitrarily localized in the bulk. 
An important resulting consequence for collider searches is that  light fermions have small couplings to KK modes (including the graviton) while the top quark and the Higgs have a large coupling to KK modes. 
%LHC probes of the RS setup depend on the mass scale of the KK modes which we now discuss.

\subsection{Mass spectrum}

\begin{figure}[tb]
\begin{center}
\includegraphics[width=10.5cm,height=9cm]{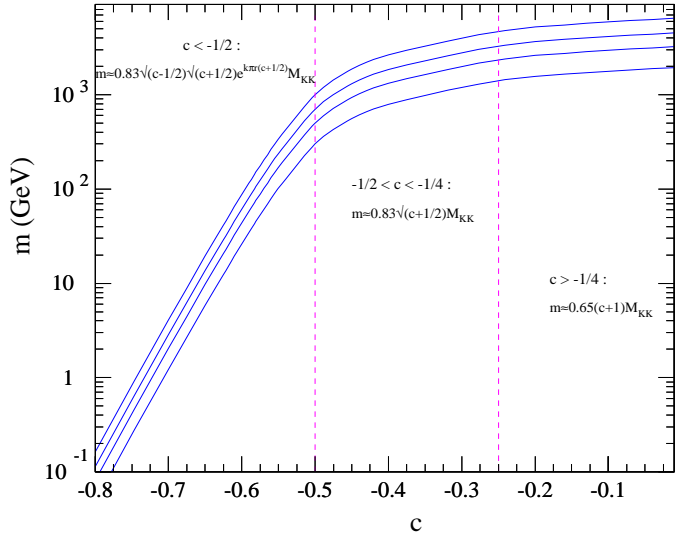}
\caption[]{Mass of the (-+) KK fermion as a function of its $c$ parameter for different values of the 
KK gauge boson mass $M_{KK}=$3, 5, 7, 10 TeV (from bottom to top). For large and negative $c$, the KK fermion can be infinitely light. 
%For KK fermions belonging to the GUT multiplet containing the RH 
%top, $c \sim -1/2$.
}
\label{LZPmass}
\end{center} 
\end{figure}

The challenging 
aspect of RS collider phenomenology is that the mass scale of KK gauge bosons is constrained to be at least a  
few TeV by the electroweak and flavor precision tests. This is 
in part due to the absence of a parity symmetry
(analogous to $R$-parity in SUSY or KK parity in UED), allowing tree-level exchanges
to contribute to the precision observables. 
Nevertheless, in contrast with UED, there is not necessarily a single KK scale and some KK fermions are allowed to have a mass significantly different from  KK gauge bosons.
Indeed, the mass spectrum of KK fermions depends strongly on the type of boundary conditions (BC) imposed at the UV and IR branes as well as on the  bulk mass parameter,  
called $c$ in Planck mass units.
The $c$ parameter also fixes the localization of the wave function of the zero modes and therefore the mass of the SM fermion. As first emphasized in 
\cite{Agashe:2004ci,Agashe:2004bm}, there can be very light KK fermions as a consequence of the  top compositeness.
BC are commonly modelled by either Neumann ($+$)
or Dirichlet ($-$) BC\footnote{For a comprehensive description
 of boundary conditions of  fermions on an interval, see \cite{Csaki:2003sh}.} in orbifold compactifications. 
 Five-dimensional fermions lead to two chiral fermions in 4D, only one 
of which gets a zero mode to reproduce the chiral Standard Model fermion. 
SM fermions are associated with ($++$) BC (first sign is for Planck brane, second
 for TeV brane).
The other chirality is ($--$) and does not have a zero mode. 
In the particular case of the breaking of the grand unified gauge group (GUT) to the SM, 
fermionic GUT partners which do not have zero modes are assigned 
Dirichlet boundary conditions on the Planck brane, i.e. they have ($-+$) boundary conditions
\footnote{Consistent extra dimensional GUT models require a replication of 
GUT multiplets as the zero-modes SM particles are obtained from different multiplets.}. 
When computing the KK spectrum of $(-+)$ fermions 
one finds that for $c < 1/2$ the lightest KK fermion 
is lighter than the lightest KK gauge boson.
 For the particular 
case  $c < -1/2$, the mass of this KK fermion is exponentially
smaller than that of the gauge KK mode.
Fig. \ref{LZPmass} shows the mass of the lightest ($-+$) KK fermion as a function of $c$ and for different values of the KK gauge boson mass $M_{KK}$.
There is an intuitive argument for the lightness of
the KK fermion:
for $c \ll 1/2$, the zero-mode of the  fermion with $(++)$ boundary condition
is localized near the TeV brane. 
Changing the boundary condition to $(-+)$
makes this ``would-be'' zero-mode massive, but since it is localized near the  TeV
brane, the effect of changing the boundary condition on the Planck
brane is suppressed, resulting in a small mass for the would-be zero-mode.

%%%%%%%%%%%%%%%%%%%%%%%%%%%%%%%%%%%%%
%%%%%%%%%%%%%%%%%%%%%%%%%%%%%%%%%%%%%
Therefore, the scale for KK fermions can be different from the scale of KK gauge bosons. 
 The lightest KK fermion is the one with the smallest $c$ parameter. 
For example, in the warped SO(10) models of Ref.~\cite{Agashe:2004ci,Agashe:2004bm},
the lightest KK fermion will come from the GUT multiplet which contains the top quark. Indeed, the top quark, being the heaviest SM fermion, is the closest to the TeV brane. This is achieved by requiring a negative $c$. Thus, all $(-+)$ KK fermions in the GUT multiplet containing the SM top quark are potentially light. Mass splittings between KK GUT partners of the top quark can have various origins, in particular due to GUT breaking in the bulk \cite{Agashe:2004ci,Agashe:2004bm}. 
Direct production at the LHC of light KK quarks leading to multi $W$ final states was studied in 
\cite{Dennis:2007tv,Contino:2008hi}.

\subsection{Collider signals}

 As a result of wave function localization, SM gauge bosons
and light fermions couple weakly to the
KK states, whereas the KK states mostly decay to top quarks and 
longitudinal $W/Z$/Higgs. 
Hence, the golden decay channels such as resonant signals of dileptons or diphotons
are suppressed. Besides, given that the KK mass scale is typically high (a few TeV) the
top quarks or $W$ or $Z$ bosons resulting from the decays of these KK states
are highly boosted.

The most widely studied particle is the 
KK gluon \cite{Agashe:2006hk,Lillie:2007yh,
Guchait:2007ux, Lillie:2007ve,Djouadi:2007eg,Guchait:2007jd,Carena:2007tn} which has the 
largest cross-section due to the QCD coupling
and decays to jet final states.
It was found that the LHC reach can be $\sim 4$ TeV, using techniques
designed specifically to identify highly boosted top quarks.

Another central prediction of the RS model are the TeV Kaluza-Klein gravitons which have 1/TeV strength couplings since their wavefunctions are peaked near the TeV brane.
They lead to distinct spin-2 resonances 
spaced according to the roots of the first Bessel function \cite{Randall:1999vf}. 
Signals from their direct production were investigated in \cite{Fitzpatrick:2007qr}.

Finally, an additional ingredient of RS phenomenology is the radion, the scalar mode of the 5D gravitational fluctuations, parameterizing the vibration mode of the inter-brane proper distance. Its mass is essentially a free parameter (it depends on the mechanism responsible for the stabilization of the inter-brane distance) but it is expected to be much lighter than the KK excitations \cite{Csaki:2000zn}.
Given that its couplings  are suppressed by the warped-down Planck scale, the radion 
 has observable effects at high energy colliders \cite{Csaki:2000zn,Hewett:2002nk, Dominici:2002jv, Dominici:2002np, Gunion:2003px,Csaki:2007ns}.
Like the Higgs, the radion is located near the IR brane and its interactions are proportional to the mass of the field it couples to (through the trace of the energy-momentum tensor).
The Higgs and radion can mix through a gravitational kinetic mixing term \cite{Giudice:2000av} with interesting consequences \cite{Csaki:2000zn,Hewett:2002nk,Dominici:2002jv, Dominici:2002np,Gunion:2003px}. Signals from direct production of the radion have been studied in 
\cite{Rizzo:2002pq,Csaki:2007ns,Toharia:2008tm}.

In addition to the searches for the radion, the KK graviton and the KK gluons, 
other studies of warped phenomenology at the LHC have dealt with 
KK neutral electroweak gauge bosons 
\cite{Agashe:2007ki}, KK (heavy) fermions \cite{Davoudiasl:2007wf}.
All these signatures are common to RS models. Besides, new characteristic 
predictions appear in more specific models. We mentioned the light KK fermions which appear for instance in warped GUT models with $Z_3$ symmetry and LZP dark matter \cite{Agashe:2004ci,Agashe:2004bm} but more generically in models where the Higgs is a pseudo-Goldstone boson \cite{Contino:2003ve,Contino:2008hi}. They are for instance predicted in the gauge-Higgs unification models of Ref.~\cite{Panico:2008bx} which contain a $Z_2$ mirror symmetry.
The associated signatures are jets and missing energy and benefit from a large cross section (vector-like quarks are pair-produced via the standard QCD interactions).
Other interesting warped models with distinctive phenomenologies have been proposed such as warped supersymmetric models \cite{Nomura:2004zs}. A recent finding and a generic prediction of 5D models is the existence of stable skyrmion configurations \cite{Pomarol:2007kr} with phenomenological consequences that remain to be investigated.

%%%%%%%%%%%%%%%%%%%%%%%%%%%%

\section{SUSY-UED discrimination at the LHC}
\label{sec:susyUED}

As discussed in Section~\ref{sec:flatUEDcol}, the
generic collider signatures of the minimal UED models 
involve jets, leptons, and missing energy, just like the
signatures of supersymmetry.
%supersymmetry signatures. %of Chapter \ref{Chap:Plehn}.
In addition, the couplings of the KK partners are equal
to those of their SM counterparts. The same property is shared
by the superpartners in SUSY models. A natural question, 
therefore, is whether 
a given UED model can be experimentally differentiated
from supersymmetry and vice versa. This issue is the subject 
of this section.

In general, there are two fundamental differences between UED and SUSY.
\begin{enumerate}
\item The spins of the SM particles and their KK partners are the same, 
while in SUSY they differ by $\frac{1}{2}$.
\item For each particle of the Standard Model, the UED 
models predict an infinite tower of new particles 
(Kaluza-Klein partners). In contrast, the simplest SUSY models 
contain only one partner per SM particle.
\end{enumerate}
Thus the best way to discriminate between UED and SUSY is 
to either measure the spins of the new particles
or to explore the higher level states of the KK tower.

Spin determinations in missing energy events are rather challenging
(especially at hadron colliders), due to the presence of at least two 
invisible particles in each event, whose energies and momenta are not measured.
Ideally, one would like to be able to reconstruct the energies and momenta
of the escaping particles, in which case the spins can be determined in
one of several ways (see Secs.~\ref{sec:discthr}, \ref{sec:discang} 
and \ref{sec:discphi}). However, when the momenta cannot be 
reliably determined, we are limited to studying only the properties
of the particles which are visible in the detector, e.g.~their
invariant mass distributions (see Sec.~\ref{sec:discmass}).

On the other hand, it is relatively easier to find the higher KK modes, 
as long as they are produced abundantly. In particular, the $n=2$
excited KK states have positive KK parity and can directly decay 
into a pair of SM particles. Such higher level KK particles can then be looked for 
via traditional resonance searches (see Sec.~\ref{sec:discres}). 
The observation of a rich resonance structure would be quite 
indicative of UED. 

\subsection{Spin measurements from invariant mass distributions}
\label{sec:discmass}

\begin{figure}[tbp]
\unitlength=1.5 pt
\SetScale{1.5}
\SetWidth{1.0}      % line    size control
\normalsize    %  letter  size control
{} \qquad\allowbreak
%  diagram # 1
\centerline{
\begin{picture}(330,100)(0,65)
\SetColor{Gray}
\Line(100,80)(120,120)
\Line(150,80)(170,120)
\Line(200,80)(220,120)
\SetColor{Red}
\Text( 75, 86)[c]{\Red{$D$}}
\Text(125, 86)[c]{\Red{$C$}}
\Text(175, 86)[c]{\Red{$B$}}
\Text(225, 86)[c]{\Red{$A$}}
\SetWidth{1.2}      % line    size control
\Line(50,80)(250,80)
%\Text(100, 55)[c]{\Black{$x=\frac{m_C^2}{m_D^2}$}}
%\Text(150, 55)[c]{\Black{$y=\frac{m_B^2}{m_C^2}$}}
%\Text(200, 55)[c]{\Black{$z=\frac{m_A^2}{m_B^2}$}}
\Text(120,129)[c]{\Black{$j$}}
\Text(170,129)[c]{\Black{$\ell^\pm$}}
\Text(220,129)[c]{\Black{$\ell^\mp$}}
\SetColor{Blue}
\Vertex(100,80){2}
\Vertex(150,80){2}
\Vertex(200,80){2}
\scriptsize
\Text(100, 73)[c]{\Blue{$c_LP_L+c_RP_R$}}
\Text(150, 73)[c]{\Blue{$b_LP_L+b_RP_R$}}
\Text(200, 73)[c]{\Blue{$a_LP_L+a_RP_R$}}
\end{picture} 
}
\caption{The typical UED or SUSY cascade decay chain under 
consideration.
At each vertex we assume the most general fermion couplings 
(see Ref.~\cite{Burns:2008cp} for the exact definitions).}
\label{fig:ABCD} 
\end{figure}
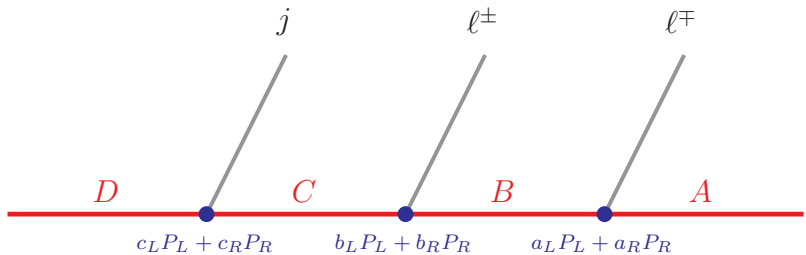

Consider the three-step decay chain exhibited in Fig.~\ref{fig:ABCD},
which is rather typical in both UED and SUSY models.
For example, in UED this chain may arise due to the
$Q_1\to Z_1 \to L_1 \to \gamma_1$ transitions 
shown in Fig.~\ref{fig:signatures}.
The measured visible decay products are a quark jet $j$ 
and two opposite sign leptons $\ell^+$ and $\ell^-$,
while the end product $A$ is invisible in the detector.
Given this limited amount of information, in principle 
there are 6 possible spin configurations for the 
heavy partners $D$, $C$, $B$ and $A$: $SFSF$, $FSFS$, $FSFV$,
$FVFS$, $FVFV$, and $SFVF$, where $S$ stands for a spin-0 scalar,
$F$ stands for a spin-$\frac{1}{2}$ fermion, and 
$V$ stands for a spin-1 vector particle.
The main goal of the invariant mass analysis here will 
be to discriminate among these 6 possibilities, and in particular
between $SFSF$ (SUSY) and $FVFV$ (minimal UED).

%%%%%%%%%%%%%%%%%%% BEGIN FIGURE %%%%%%%%%%%%%%%%%%%
%\begin{center}
%\end{center}
%%%%%%%%%%%%% END OF FIGURE ################

It is well known that the invariant mass distributions
of the visible particles (the jet and two leptons in our case)
already contain some information about the spins of the 
intermediate heavy particles $A$, $B$, $C$ and $D$ 
\cite{Barr:2004ze,Smillie:2005ar,Athanasiou:2006ef,Wang:2006hk}.
Unfortunately, the invariant mass distributions are also
affected by a number of additional factors, which have 
nothing to do with spins, such as: 
the chirality of the couplings at each vertex \cite{Kilic:2007zk,Burns:2008cp};
the fraction of events $f$ in which the cascade 
is initiated by a particle $D$ rather than its antiparticle $\bar{D}$
\cite{Barr:2004ze}; and finally, 
the mass splittings among the heavy partners \cite{Smillie:2005ar}.
Therefore, in order to do a pure and model-independent 
spin measurement, one has to somehow eliminate the effect 
of those three extraneous factors.

Fortunately, the masses of $A$, $B$, $C$ and $D$ can 
be completely determined ahead of time, for example by measuring the kinematic 
endpoints of various invariant mass distributions made out of 
the visible decay products in the decay chain of Fig.~\ref{fig:ABCD}
\cite{Hinchliffe:1996iu,Allanach:2000kt,Gjelsten:2004ki,Gjelsten:2005aw},
or through a sufficient number of transverse mass
measurements \cite{Lester:1999tx,Burns:2008va}.
Once the mass spectrum is thus determined, we are still left with
a complete lack of knowledge regarding the 
coupling chiralities and particle fraction $f$.
In spite of this residual ambiguity, the spins can 
nevertheless be determined, at least as a matter of principle 
\cite{Burns:2008cp}. To this end, one should not make any
a priori assumptions and instead consider 
the most general fermion couplings at each vertex in Fig.~\ref{fig:ABCD}
and any allowed value for the parameter $f$.
Then, the invariant mass distributions should be used to make 
separate independent measurements of the spins, on one hand, and 
of the couplings and $f$ fraction, on the other.
Following the analysis of \cite{Burns:2008cp}, here
we shall illustrate this procedure with 
two examples --- one from supersymmetry and one from UED.
The corresponding mass spectra, chirality parameters and 
particle-antiparticle fraction $f$ for each case
are listed in Table \ref{tab:points}.

\begin{table}
\centerline{
\begin{tabular}{|c|c|c|}
\hline 
                              & SPS1a                 & UED500\\
\hline
$(m_A, m_B, m_C, m_D)$ in GeV  & (96, 143, 177, 537)   & (501, 515, 536, 598)\\
$(f, \bar{f})$                & (0.7, 0.3)            & (0.66, 0.34)\\
$(a_L, a_R, b_L, b_R, c_L, c_R)$   & (0, 1, 0, 1, 1, 0)    & (1, 0, 1, 0, 1, 0)\\
\hline
\end{tabular}
}
\caption{Two study points from SUSY (SPS1a) and UED (UED500),
characterized by particle masses, chirality coefficients 
and particle-antiparticle fractions $f$ and $\bar{f}$. 
\label{tab:points}
}
\end{table}

\begin{figure}[htbp]
\hspace*{0.4cm}
\includegraphics[width=6.cm]{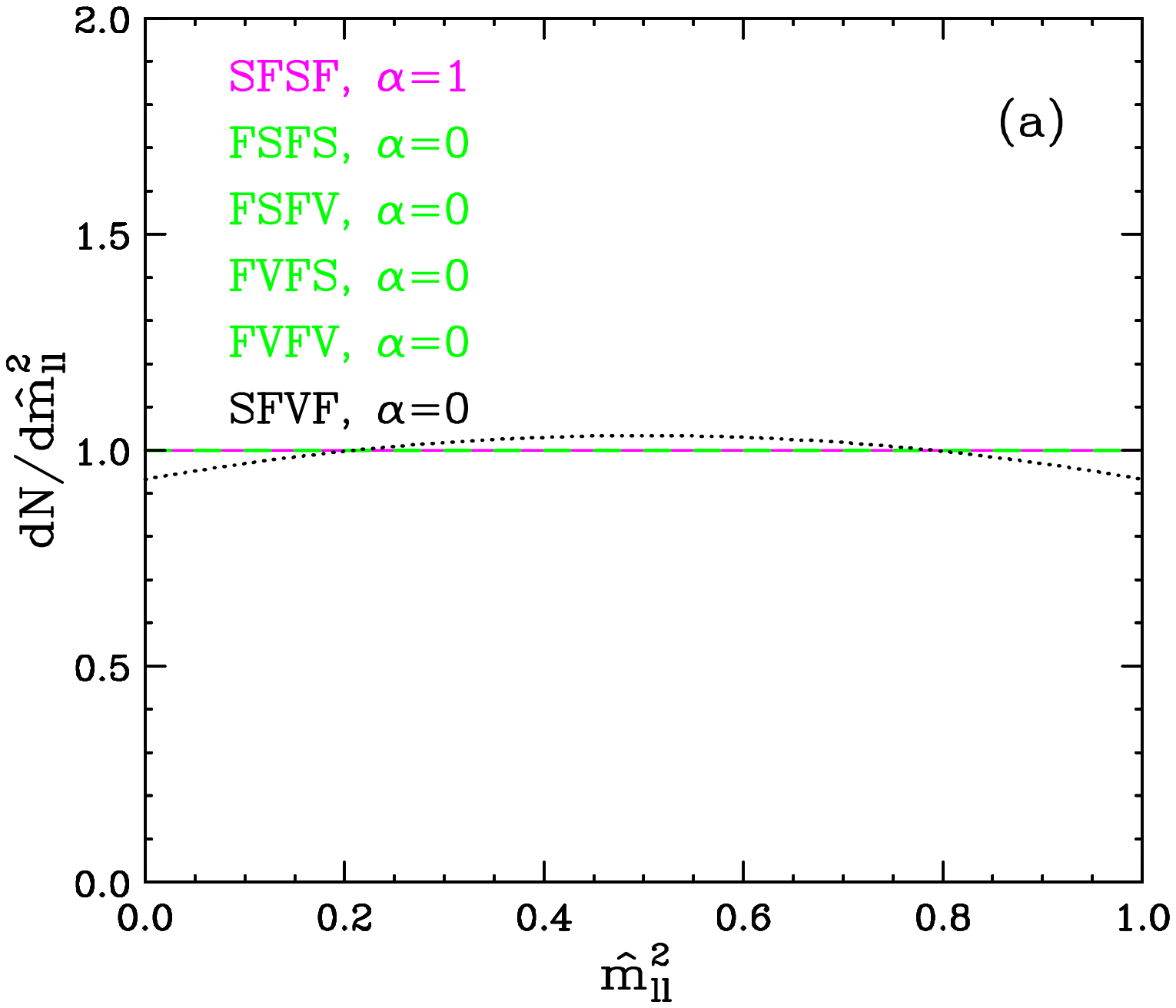} 
\includegraphics[width=6.cm]{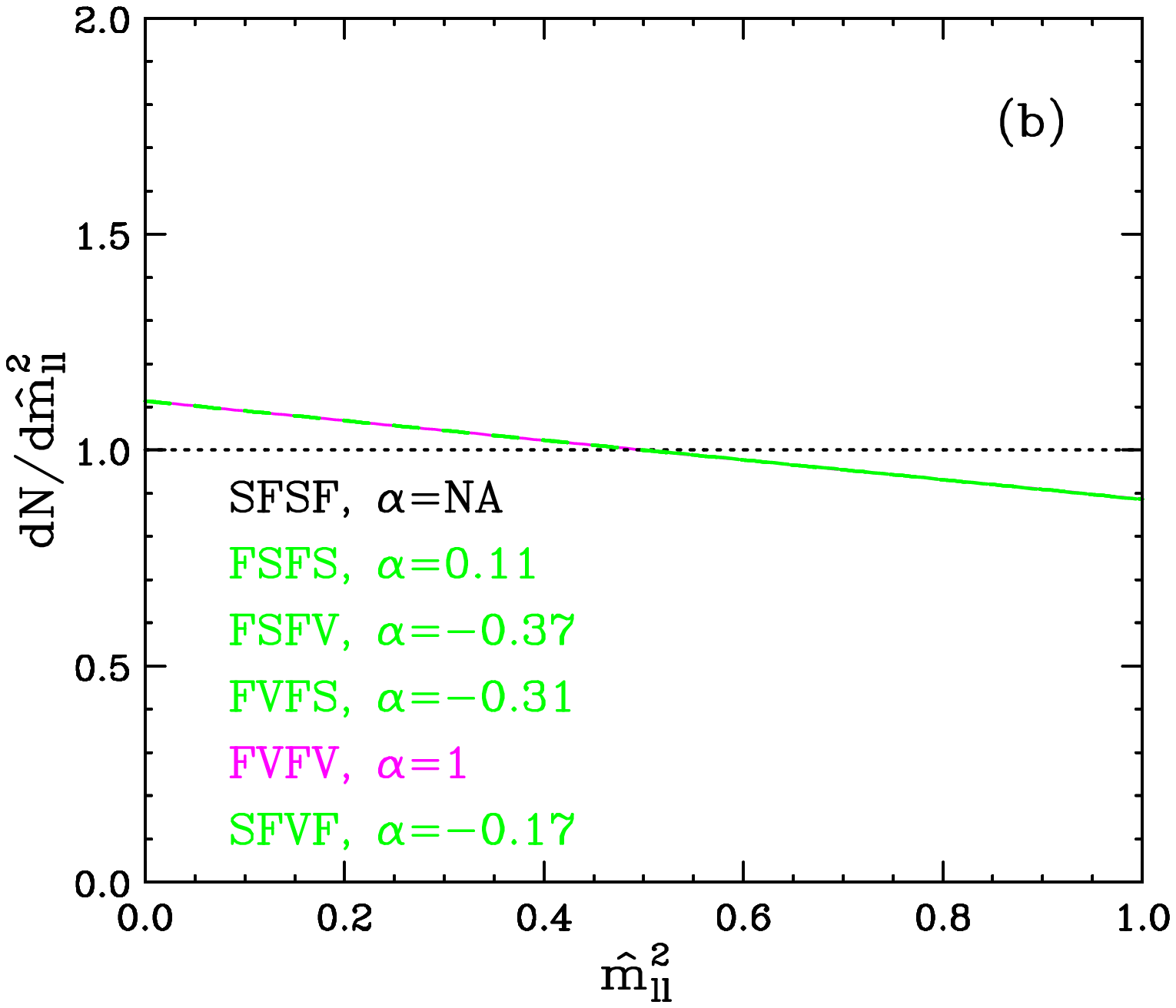}
\\
\hspace*{0.4cm} 
\includegraphics[width=6.cm]{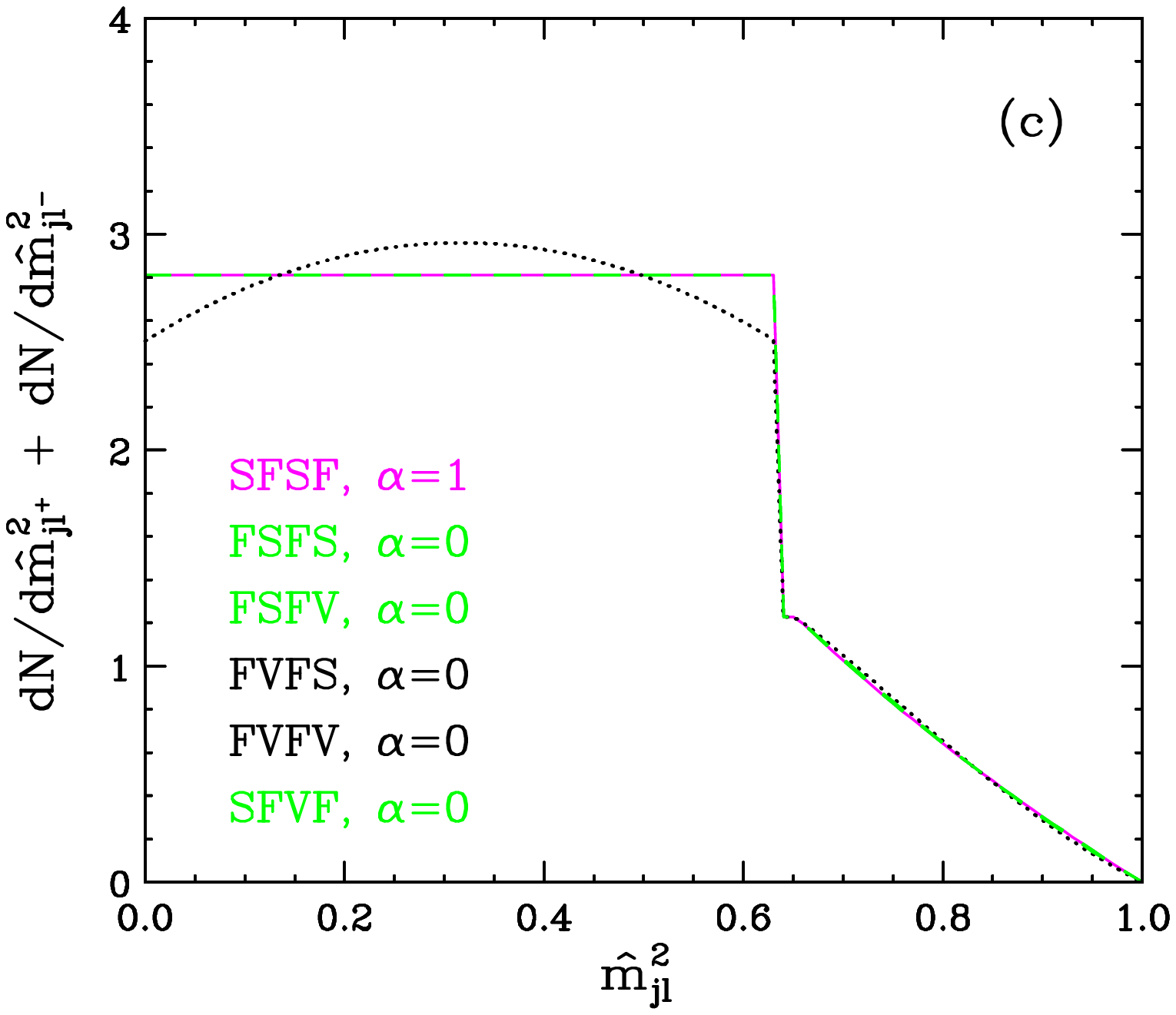} 
\includegraphics[width=6.cm]{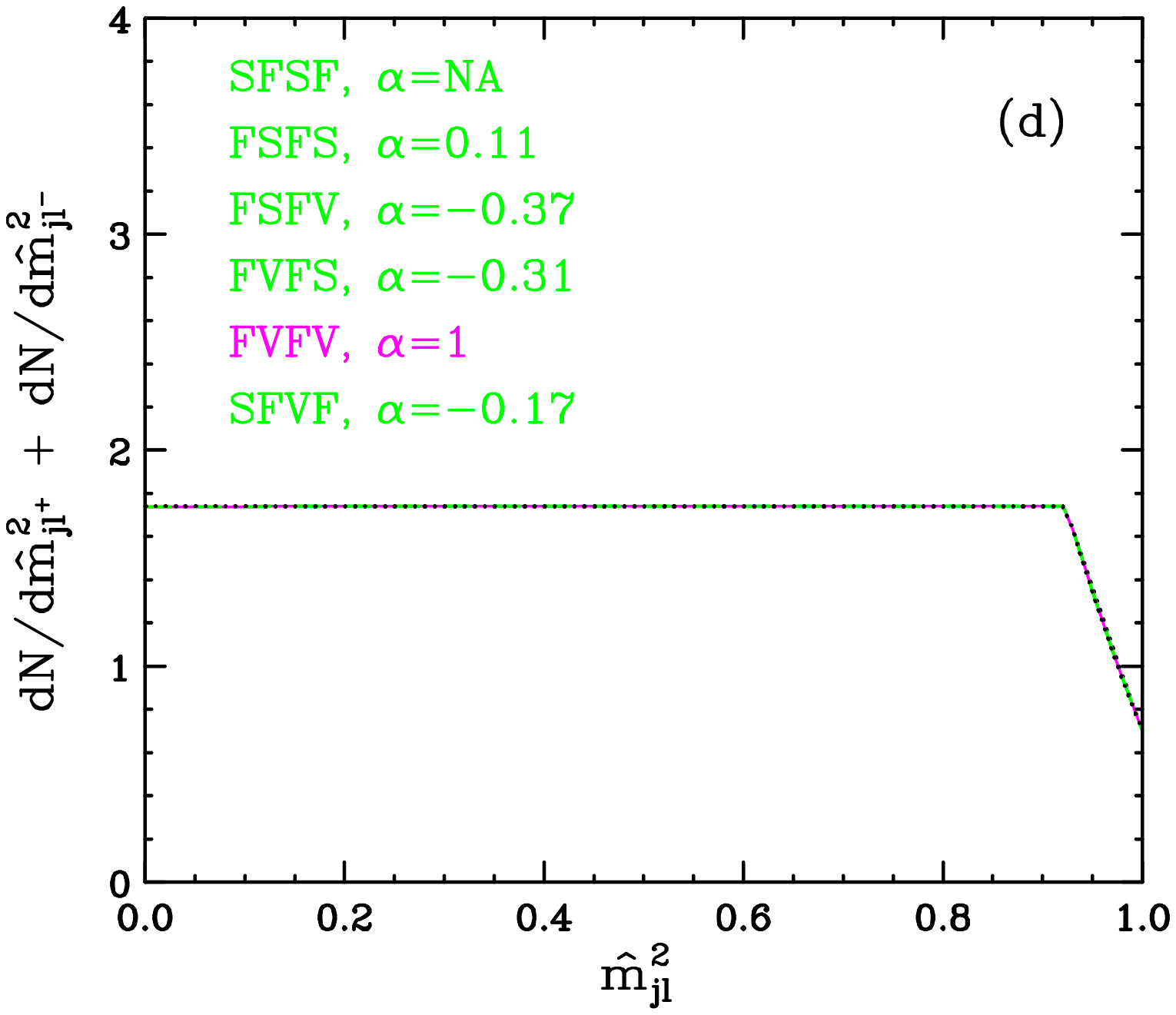}
\\
\hspace*{0.4cm}
\includegraphics[width=6.cm]{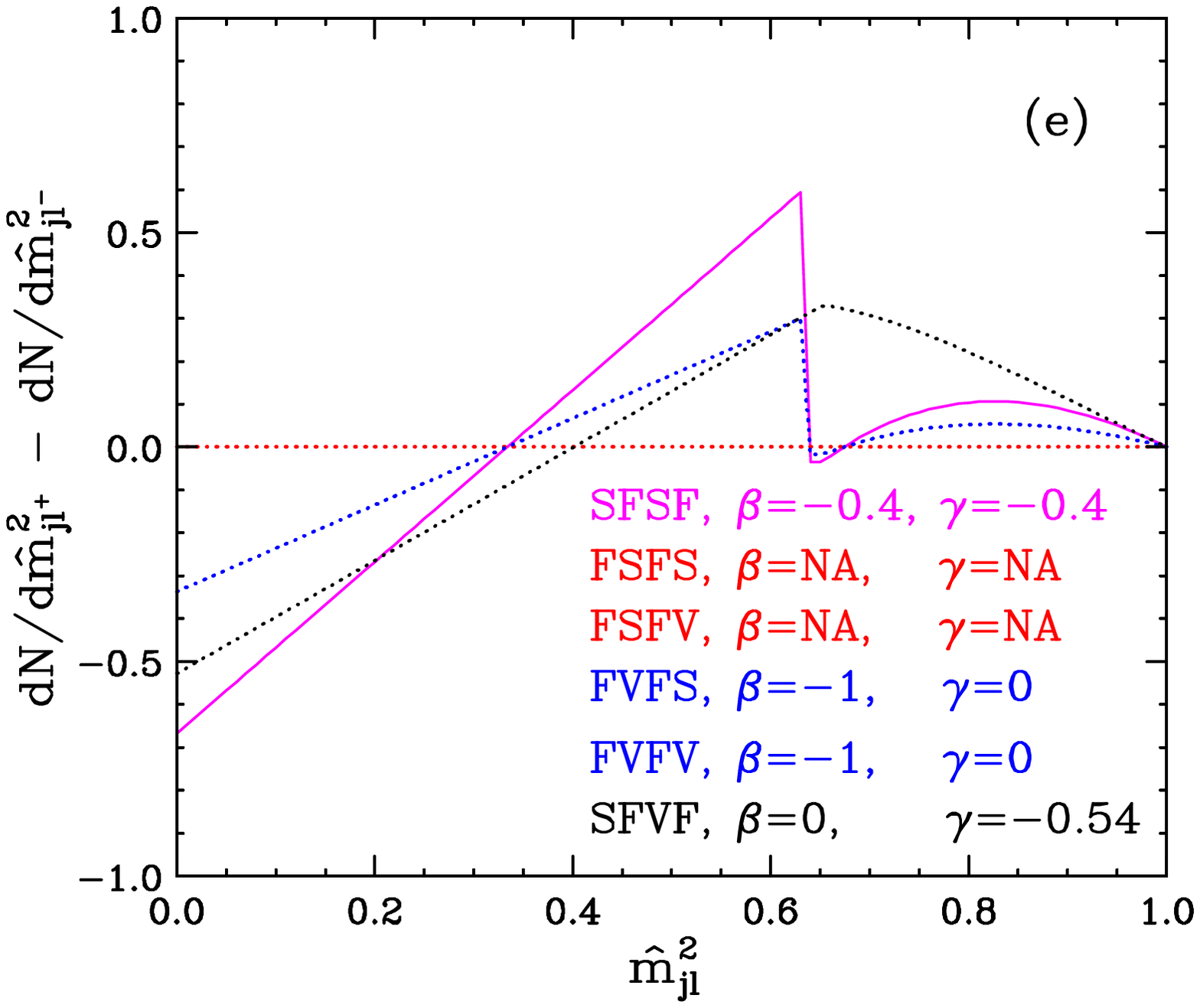} 
\includegraphics[width=6.cm]{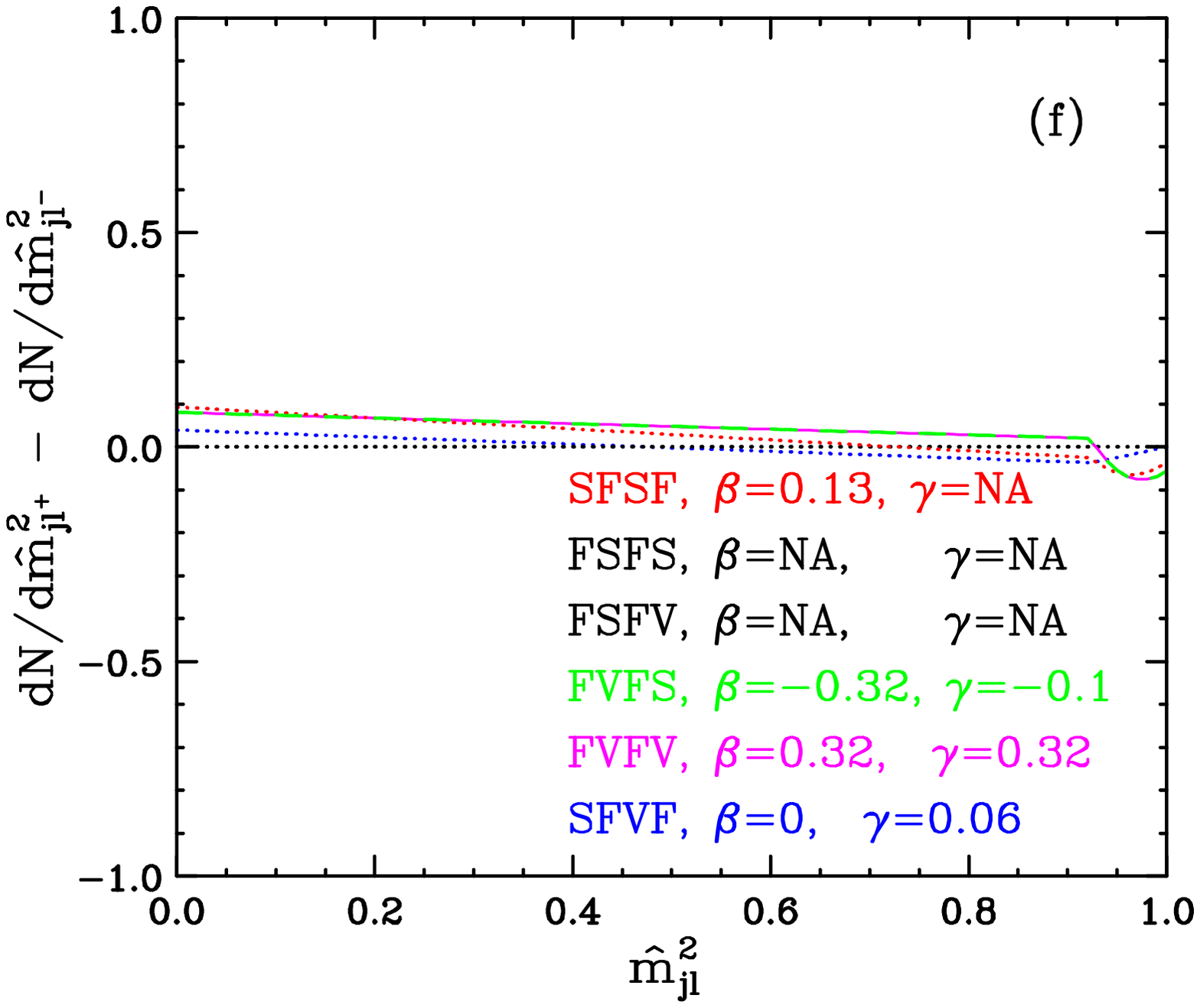}
\caption{Best fits to the three invariant mass distributions 
predicted for the SUSY (a,c,e) and UED (b,d,f) study points
from Table~\ref{tab:points}.
The solid (magenta) line in each plot represents the input 
``data'' while the other (dotted or dashed) lines are 
the best fits to the data for each of the remaining 5 
spin configurations. Dashed green lines indicate a perfect 
match to the data, while dotted lines of any other color 
imply that the best fit fails to perfectly reproduce the data. 
The best fit values for the relevant coefficients 
($\alpha$, $\beta$ and $\gamma$) are also shown.
\label{fig:invmass}
}
\end{figure}

Given the three visible particles from the decay chain of Fig.~\ref{fig:ABCD},
one can form three well-defined two-particle invariant mass distributions: 
one dilepton ($\ell^+\ell^-$), and two jet-lepton ($j\ell^+$ and $j\ell^-$) distributions.
For the purposes of the spin analysis, it is actually more convenient to consider 
the sum and the difference of the two jet-lepton distributions \cite{Burns:2008cp}.
The shapes of the resulting invariant mass distributions are given
schematically by the following formulas \cite{Burns:2008cp}:
\begin{eqnarray}
\left( \frac{dN}{dm^2_{\ell\ell}}\right)_S &=&
F_{S;\delta}^{(\ell\ell)}(m^2_{\ell\ell})
+ \alpha\, F_{S;\alpha}^{(\ell\ell)}(m^2_{\ell\ell}) 
\label{L+-} \\ [2mm]
\left( \frac{dN}{dm^2_{j\ell^+}}\right)_S + 
\left( \frac{dN}{dm^2_{j\ell^-}}\right)_S &=&
F_{S;\delta}^{(j\ell)}(m_{j\ell}^2)
+ \alpha\, F_{S;\alpha}^{(j\ell)}(m_{j\ell}^2)  
\label{S+-} \\ [2mm]
\left( \frac{dN}{dm^2_{j\ell^+}}\right)_S -
\left( \frac{dN}{dm^2_{j\ell^-}}\right)_S &=&
 \beta \, F_{S;\beta }^{(j\ell)}(m_{j\ell}^2) 
+\gamma\, F_{S;\gamma}^{(j\ell)}(m_{j\ell}^2)\ ,
\label{D+-}
\end{eqnarray}
where the functions $F$, given explicitly in \cite{Burns:2008cp},
are known functions of the masses of particles $A$, $B$, $C$ and $D$. 
As indicated by the 
index $S$, there is a separate set of $F$ functions for each
spin configuration: 
%
%\begin{equation}
$S=\{SFSF, FSFS, \\ FSFV,FVFS, FVFV, SFVF \} \, .$
%\end{equation}
% 
Thus the functions $F$ contain the 
pure spin information. On the other hand, the coefficients
$\alpha$, $\beta$ and $\gamma$ encode 
all of the residual model dependence, namely
the effect of the coupling chiralities and 
particle-antiparticle fraction $f$. Since 
the coefficients $\alpha$, $\beta$ and $\gamma$ are a priori 
unknown, they will need to be determined from experiment, 
by fitting the predicted shapes (\ref{L+-}-\ref{D+-}) to the data.
The results from this exercise for the two study points from Table 
\ref{tab:points} are shown in Fig.~\ref{fig:invmass}.
The solid (magenta) lines in each
panel represent the input invariant mass distributions
which will be presumably measured by experiment.
The other (dotted or dashed) lines are the best fits to this
data, for each of the remaining 5 spin configurations $S$. 
The color code is the following. A dashed (green) line
indicates that the trial model fits the input data perfectly,
while a dotted line implies that the fit fails to match 
the input data. The best fit values for the relevant coefficients 
($\alpha$, $\beta$ and $\gamma$) for each case are also shown,
except for those cases (labeled by ``NA'') where they
are left undetermined by the fit. Dotted lines of the same 
color imply that they are identical to each other, 
yet different from the input ``data''.

%
%\begin{center}
%\end{center}
%

The results from Fig.~\ref{fig:invmass} show that 
the success of the spin measurement method depends on the
type of new physics which happens to be discovered.
In the case of supersymmetry (panels a,c,e), 
the spin chain can be unambiguously determined 
to be $SFSF$. Furthermore, this can be done 
solely on the basis of the distribution (\ref{D+-}), 
which is sufficiently powerful to rule out all of 
the remaining 5 spin chain candidates\footnote{The distribution
(\ref{D+-}) is closely related to the lepton charge 
asymmetry proposed in \cite{Barr:2004ze}.}. 
On the other hand, the UED case (panels b,d,f)
is more challenging, and the end result is inconclusive --- 
both models $FVFS$ and $FVFV$ are able to perfectly fit all three
distributions (\ref{L+-}-\ref{D+-}). This confusion 
is not related to the specific choice of our UED study point,
but is in fact a general feature of any $FVFV$ and $FVFS$ 
(as well as $FSFV$ and $FSFS$) pair of models \cite{Burns:2008cp}.
In summary, it appears that through studies of the {\em shapes} 
of the invariant mass distributions of the visible decay products
in a chain such as the one in Fig~\ref{fig:ABCD}, one should
be able to discriminate between SUSY and UED, while the 
specific type of UED model may remain uncertain.

\subsection{Higher level KK resonance searches}
\label{sec:discres}

\begin{figure}
\centerline{
\includegraphics[width=6.5cm]{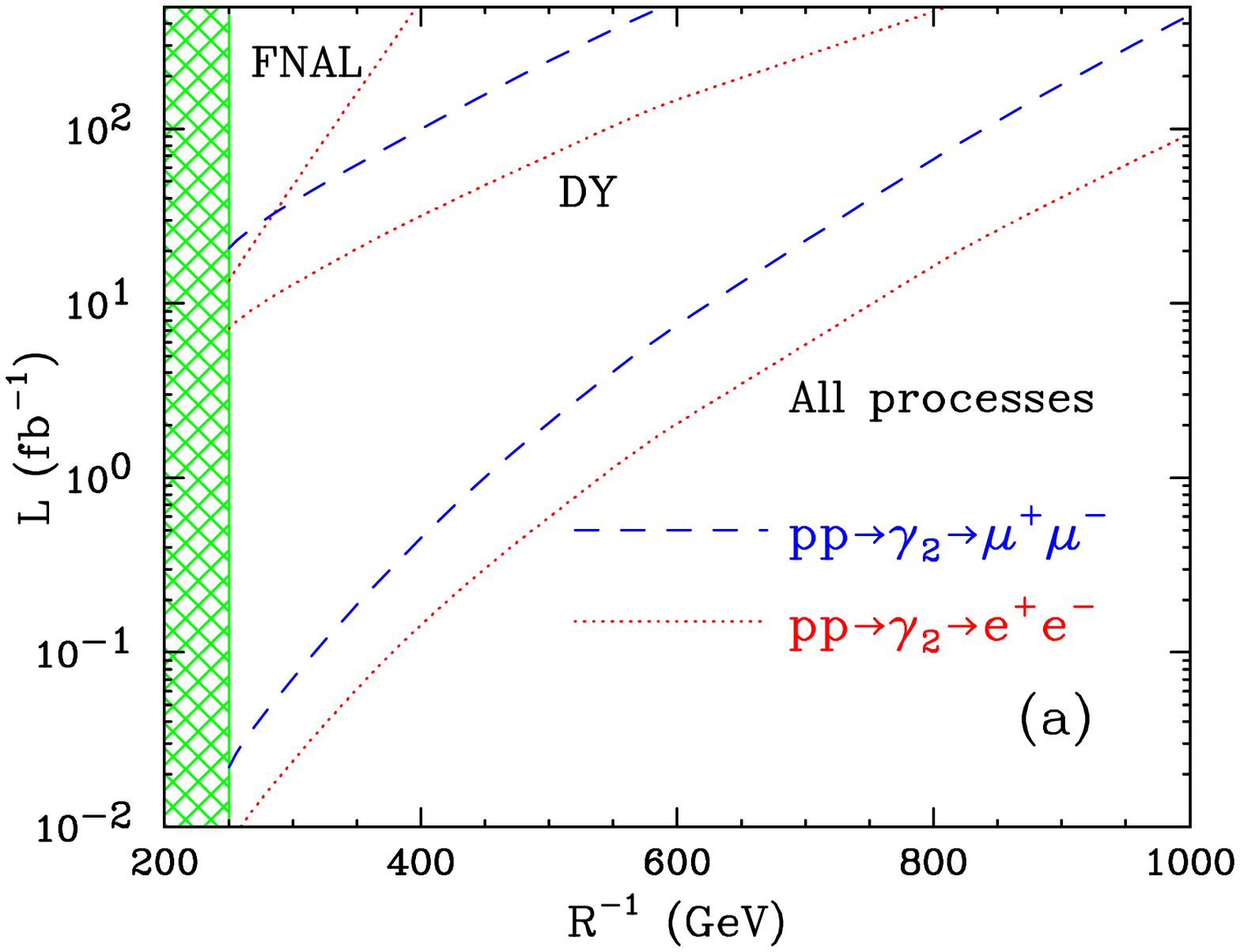}\hspace{0.3cm}
\includegraphics[width=6.5cm]{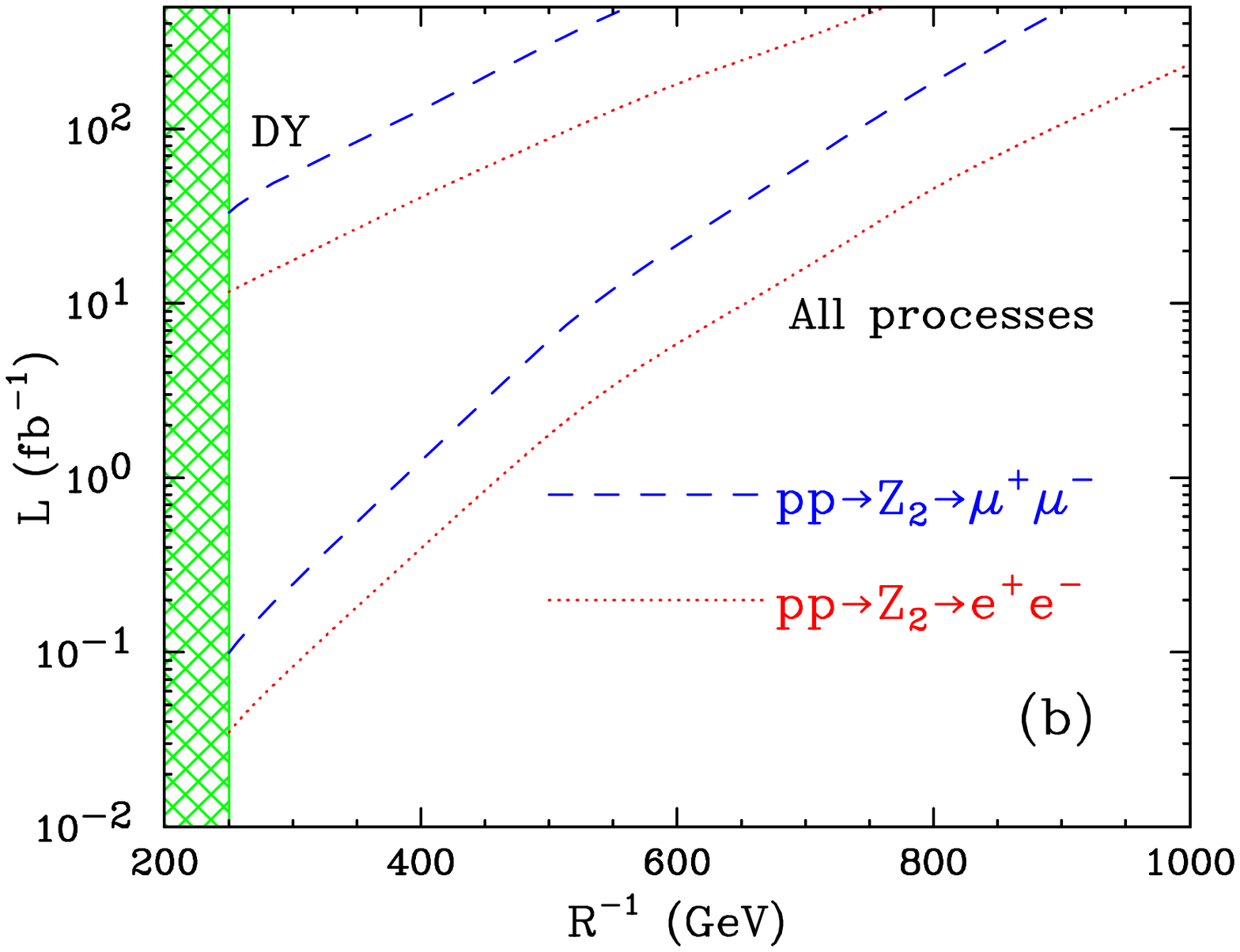} }
\caption{Total integrated luminosity needed for a
5$\sigma$ discovery of (a) $\gamma_2$ and (b) $Z_2$
as a dielectron (red) or dimuon (blue) resonance.
The shaded area below $R^{-1}$ = 250 GeV indicates the region
disfavored by precision electroweak data. 
Figures taken from Ref. \cite{Datta:2005zs}.}
\label{fig:reach}
\end{figure}

As already mentioned earlier, the discovery of the higher level 
KK resonances would be another strong indication of the UED scenario.
At hadron colliders like the LHC, resonance searches are easiest
in the dilepton (dimuon or dielectron) channels. The corresponding 
5$\sigma$ discovery reach for (a) $\gamma_2$ and (b) $Z_2$
is shown in Fig.~\ref{fig:reach} \cite{Datta:2005zs}.
In each plot, the upper set of lines labeled ``DY''
makes use of the single $\gamma_2$ or $Z_2$ production only, 
while the lower set of lines (labeled ``all processes'') 
includes in addition indirect $\gamma_2$ and $Z_2$ production 
from $n = 2$ KK quark decays. The red dotted
line marked ``FNAL'' in the upper left corner of (a) 
reflects the expectations for a $\gamma_2 \to e^+e^-$
discovery at the Tevatron in Run II.

% Level 2 resonances \cite{Datta:2005zs}
%

\begin{figure}
\centerline{
\includegraphics[width=6.5cm]{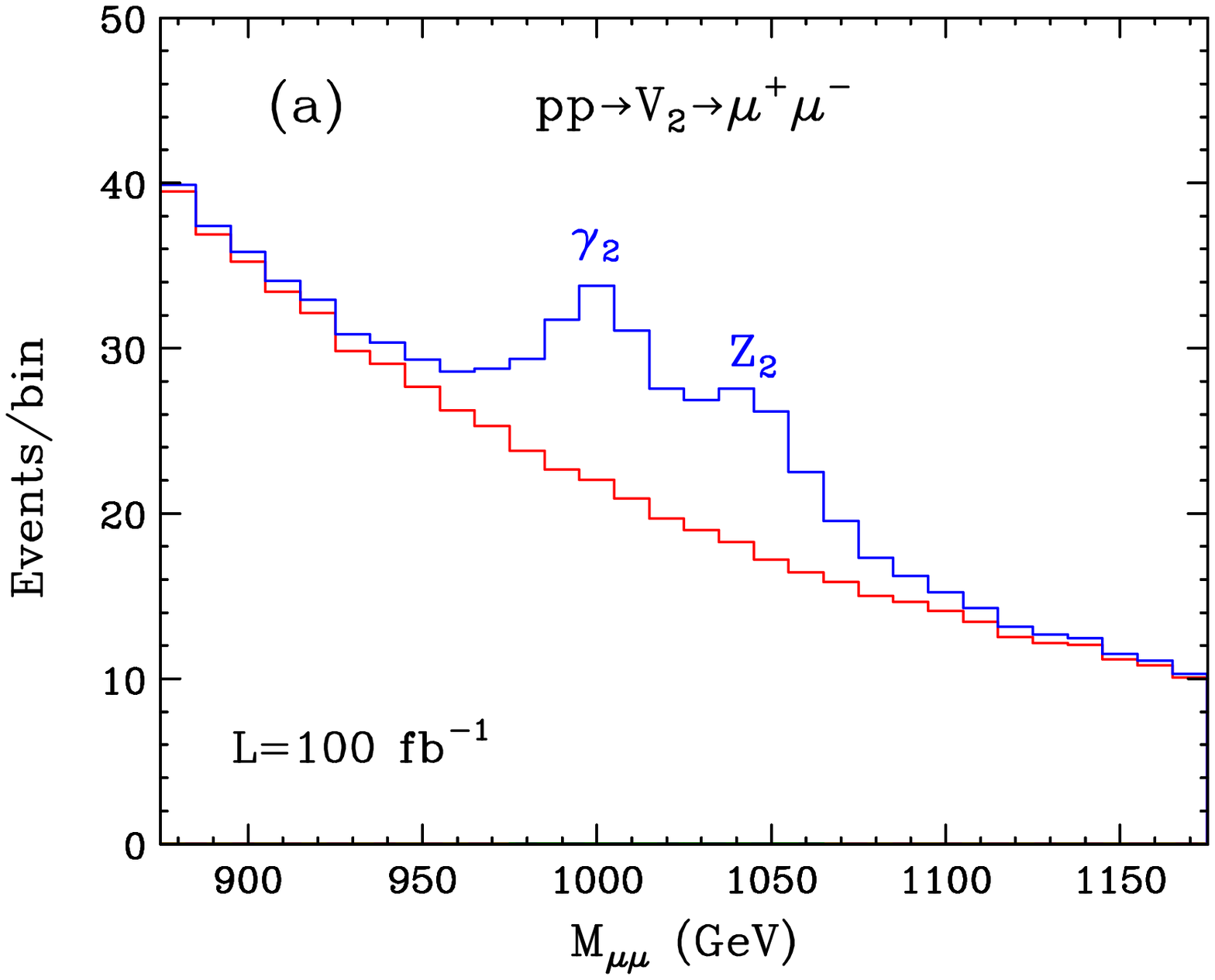} \hspace{0.3cm}
\includegraphics[width=6.5cm]{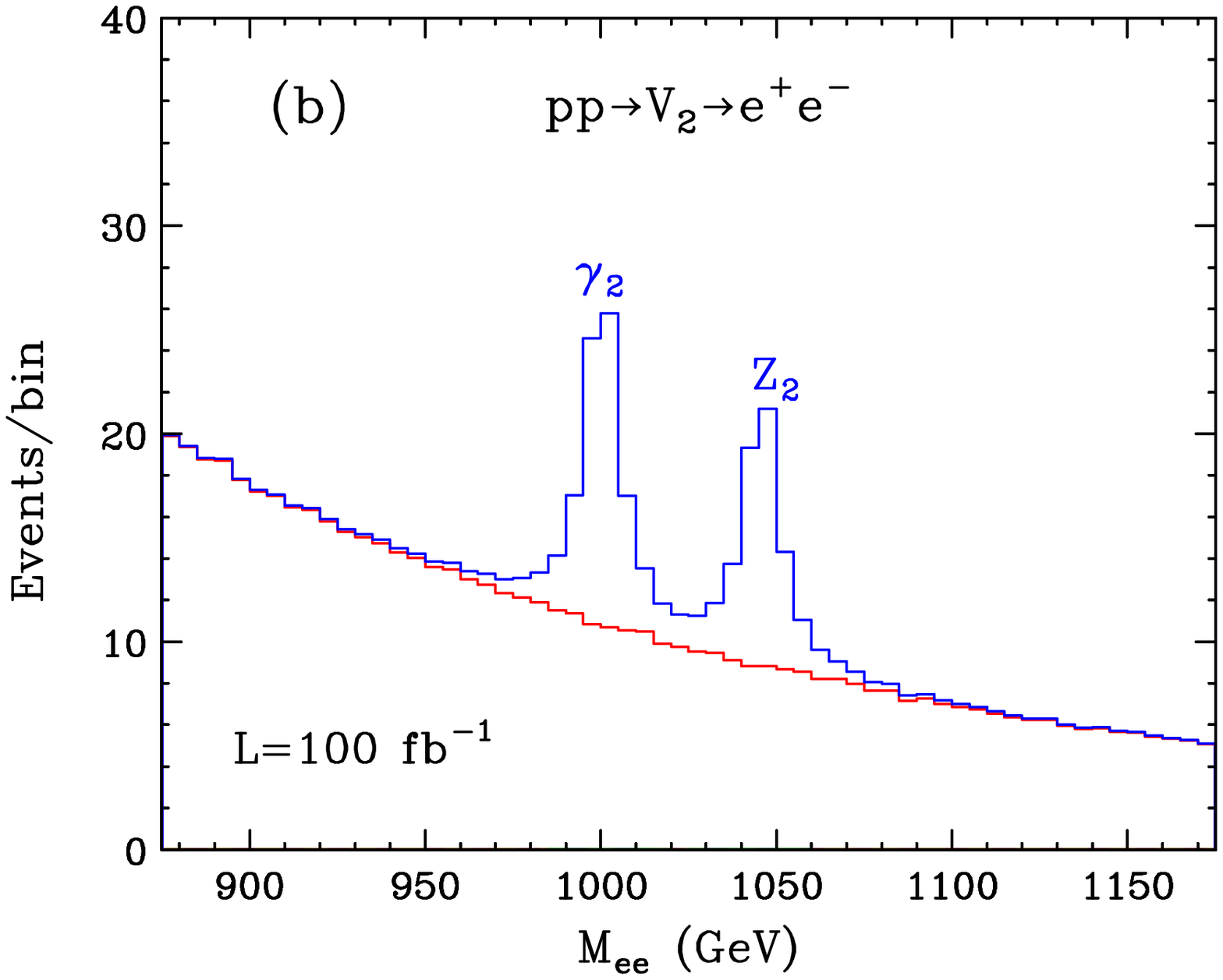} }
\caption{
The $V_2 \equiv \gamma_2 , Z_2$ diresonance structure in UED with $R^{-1}$= 500 GeV, 
for the (a) dimuon and (b) dielectron channel at the LHC with L = 100 fb$^{-1}$. 
The SM background is shown with the (red) continuous underlying histogram. 
Figures taken from Ref. \cite{Datta:2005zs}.}
\label{fig:two_reso}
\end{figure}

While the discovery of an $n = 2$ KK gauge bosons would be a strong
argument in favor of UED, any such resonance by itself is not a 
sufficient proof, since it resembles an ordinary $Z^\prime$ gauge
boson in supersymmetry. An important corroborating evidence in 
favor of UED would be the simultaneous discovery of several, 
{\em rather degenerate} KK gauge boson resonances, for which 
there would be no good motivation in generic SUSY models.
The crucial question therefore is whether one can resolve the 
different $n = 2$ KK gauge bosons as individual resonances. 
For this purpose, one would need to see a double peak structure 
in the invariant mass distributions, as illustrated in Fig.~\ref{fig:two_reso}.
We see that the diresonance structure is easier to detect in the dielectron
channel, due to the better mass resolution. In dimuons, with L = 100 fb$^{-1}$ 
the structure is also beginning to emerge. We should note that 
initially the two resonances will not be separately distinguishable, 
and each will in principle contribute to the discovery of a broad bump.
In this sense, the reach plots in Fig.~\ref{fig:reach} 
are rather conservative, since they do not combine 
the two signals from $Z_2$ and $\gamma_2$, but show the 
reach for each resonance separately.

\subsection{Spin measurements from production cross-sections}
\label{sec:discthr}

The spin of the new particles can also be inferred from the 
threshold behavior of their production cross-section
\cite{Battaglia:2005ma,Battaglia:2005zf}.
For an $s$-channel diagram mediated by a gauge boson, 
the pair production cross-section for a spin-0 particle 
behaves like $\sigma \sim \beta^3$ 
while the cross-section for a spin-$\frac{1}{2}$ 
particle behaves as $\sigma \sim \beta$, 
where $\beta = \sqrt{1 - \frac{4m^2}{s}}$, and 
$\sqrt{s}$ is the total center-of-mass energy, while $m$ is the 
mass of the new particle. At lepton colliders the threshold behavior 
can be easily studied by varying the beam energy and
measuring the corresponding total cross-section, without any need 
for reconstructing the kinematics of the missing particles.
In contrast, at hadron colliders the initial state partons
cannot be controlled, so in order to apply this method, one has to  
fully reconstruct the final state, which is rather difficult
when there are two or more missing particles.

The total production cross-section may also be used as an indicator
of spin \cite{Datta:2005vx,Kane:2008kw}. For example, 
the total cross-sections of the fermion KK modes in UED are 
5-10 times larger than the corresponding cross-sections for
scalar superpartners of the same mass. However, the measurement
of the total cross-section necessarily involves additional 
model-dependent assumptions regarding the branching fractions, 
the production mechanism, etc.

\subsection{Spin measurements from angular distributions}
\label{sec:discang}

Perhaps the most direct indication of the spin of the new particles is 
provided by the azimuthal angular distribution at production 
\cite{Battaglia:2005ma,Battaglia:2005zf}.
Assuming production through an $s$-channel gauge boson,
the angular distribution for a spin-0 particle is $\sim (1 - \cos^2\theta)$,
where $\theta$ is the azimuthal production angle in the center-of-mass frame.
In contrast, the distribution for a spin-$\frac{1}{2}$ particle 
is $\sim (1 + \cos^2\theta)$.
Unfortunately, reconstructing the angle $\theta$ 
generally requires a good knowledge of the momentum 
of the missing particles, which is only possible at a lepton collider.
Applying similar ideas at the LHC, one finds 
that typically quite large luminosities are needed \cite{Barr:2005dz}.

\subsection{Spin measurements from quantum interference}
\label{sec:discphi}

When a particle is involved in both the production and the decay, its spin $s$
can also be inferred from the angle $\phi$ between the production and decay planes
\cite{Buckley:2008eb,Buckley:2008pp,Buckley:2007th}.
The cross section can be written as 
\begin{equation}
\frac{d\sigma}{d \cos\phi} = a_0 + a_1 \cos\phi + a_2 \cos 2\phi + \cdots + a_{2s} \cos 2s \phi\ .
\end{equation}
By measuring the coefficient $a_{2s}$ of the highest $\cos$ mode, 
one can in principle extract the spin $s$ of the particle.
This method is especially useful since it does not rely on the particular
production mechanism, and is equally applicable to $s$-channel and $t$-channel processes.
However, its drawback is that the $\phi$ dependence results from
integrating out all other degrees of freedom, which often leads 
to a vanishing coefficient as a result of cancellations, 
for instance, in the case of a purely vector-like coupling, or in the case of 
a $pp$ collider like the LHC.
As a result, the practical applicability of the method is rather model-dependent.

\subsection*{Acknowledgements}
KK is supported in part by the DOE under contract DE-AC02-76SF00515 and 
DE-AC02-07CH11359.
KM is supported in part by a US Department of Energy grant DE-FG02-97ER41029. 
The work of GS is supported by the European Research Council.

%%%%%%%%%%%%%%%%%%%%%%%%%%%%%%%%%%%%%%%%%%%%%%%%%%%%%%%%%%%%%%%%%%%%%%%%%%%%%%%
%%%%%%%%%%%%%%%%%%%%%%%%%%%%%%%%%%%%%%%%%%%%%%%%%%%%%%%%%%%%%%%%%%%%%%%%%%%%%%%

\end{document}